\documentclass[conference]{IEEEtran}
\usepackage{cite}

\ifCLASSINFOpdf
\else
\fi
\usepackage{amsmath}
\usepackage{xspace}
\usepackage{xparse}
\usepackage{paralist}
\usepackage{amsfonts}
\usepackage{mathtools}
\usepackage{ulem}
\usepackage{color}
\usepackage{xspace}
\usepackage[noend]{algpseudocode}
\usepackage{hyperref}
\usepackage{authblk}
\usepackage[skip=3pt,font=footnotesize]{subcaption}
\usepackage{wrapfig}

\algrenewcommand\algorithmicindent{1em}
\makeatletter
\renewcommand{\ALG@beginalgorithmic}{\footnotesize}
\makeatother

\newcommand{\ourname}{sAVSS\xspace} 
\newcommand{\ournameexplained}{scalable AVSS\xspace}

\newcommand{\mkr}[1]{\textbf{(MKR: {#1})}\xspace}
\newcommand{\alin}[1]{\textbf{(Alin: {#1})}\xspace}

\newcommand{\secref}[1]{Section~\ref{#1}\xspace}
\newcommand{\appref}[1]{Appendix~\ref{#1}\xspace}
\newcommand{\figref}[1]{Figure~\ref{#1}\xspace}
\newcommand{\prob}[1]{\ensuremath{\mathbb{P}\left({#1}\right)}\xspace}
\newcommand{\nats}[1]{\ensuremath{[{#1}]}\xspace}
\newcommand{\residues}[1]{\ensuremath{\mathbb{Z}_{#1}}\xspace}
\newcommand{\residuesNonzero}[1]{\ensuremath{\mathbb{Z}_{#1}^{\ast}}\xspace}
\newcommand{\polynomials}[1]{\ensuremath{\mathbb{Z}_{#1}[x]}\xspace}
\newcommand{\thresh}{\ensuremath{k}\xspace}
\newcommand{\numreplicas}{\ensuremath{n}\xspace}
\newcommand{\numfaulty}{\ensuremath{f}\xspace}
\newcommand{\badfraction}{\ensuremath{\ell}\xspace}
\newcommand{\coeffIdx}{\ensuremath{j}\xspace}
\newcommand{\replicaIdx}{\ensuremath{i}\xspace}
\newcommand{\replicaIdxAlt}{\ensuremath{\alt{i}}\xspace}

\newcommand{\someset}{\ensuremath{S}\xspace}
\newcommand{\cardinality}[1]{\ensuremath{|#1|}\xspace}
\newcommand{\getsr}{\;\stackrel{\$}{\leftarrow}\;}
\newcommand{\nonce}{\ensuremath{r}\xspace}

\NewDocumentCommand{\replicaIdxSubset}{ g }{\ensuremath{I\IfNoValueF{#1}{_{#1}}}\xspace}
\NewDocumentCommand{\replicaIdxSubsetAlt}{ g }{\ensuremath{\alt{I}\IfNoValueF{#1}{_{#1}}}\xspace}

\newcommand{\vssModulusQRs}{\ensuremath{q'}\xspace}
\newcommand{\qr}[1]{\ensuremath{\mathit{QR}({#1})}\xspace}
\newcommand{\dprfHash}{\ensuremath{H}\xspace}
\newcommand{\dprfZKPHash}{\ensuremath{H'}\xspace}
\newcommand{\dprfExp}{\ensuremath{\alpha}\xspace}
\newcommand{\dprfExpShare}[1]{\ensuremath{\alpha_{#1}}\xspace}
\newcommand{\dprfChallenge}[1]{\ensuremath{c_{#1}}\xspace}
\newcommand{\dprfRandom}{\ensuremath{r}\xspace}
\newcommand{\dprfContribIdx}[1]{\ensuremath{f_{#1}(\dprfInput)}\xspace}
\newcommand{\dprfProof}[1]{\ensuremath{z_{#1}}\xspace}

\newcommand{\vssGen}{\ensuremath{g}\xspace}

\newcommand{\vssGenTwo}{\ensuremath{h}\xspace}

\newcommand{\group}[1]{\ensuremath{\mathbb{G}_{#1}\xspace}}
\newcommand{\groupElem}[1]{\ensuremath{h}\xspace}
\newcommand{\polyX}{\ensuremath{x}\xspace}

\NewDocumentCommand{\blmap}{ g g }{\ensuremath{e\IfNoValueF{#1}{(#1, #2)}\xspace}}
\NewDocumentCommand{\bltrap}{ g }{\ensuremath{\tau\IfNoValueF{#1}{^{#1}}\xspace}}

\newcommand{\alt}[1]{\ensuremath{\hat{#1}}}
\newcommand{\altalt}[1]{\ensuremath{\check{#1}}}

\newcommand{\dprfScheme}{\ensuremath{\mathcal{F}}\xspace}
\newcommand{\dprfInput}{\ensuremath{x}\xspace}
\newcommand{\dprfOutput}{\ensuremath{y}\xspace}
\newcommand{\dprfDomain}{\ensuremath{\mathbf{D}}\xspace}
\newcommand{\dprfRange}{\ensuremath{\mathbf{R}}\xspace}
\newcommand{\dprfPubKey}[1]{\ensuremath{\mathit{dpk}_{#1}}\xspace}
\newcommand{\dprfPrivKey}[1]{\ensuremath{\mathit{dsk}_{#1}}\xspace}
\newcommand{\dprfSecParam}{\ensuremath{\kappa}\xspace}
\newcommand{\dprfFailure}{\ensuremath{\bot}\xspace}
\newcommand{\dprfAdversary}[1]{\ensuremath{\mathcal{A}_{#1}}\xspace}
\newcommand{\dprfCompromiseCmd}{\ensuremath{\mathit{compromise}}\xspace}

\NewDocumentCommand{\dprfInit}{ g g g g g }{\ensuremath{\mathsf{dprfInit}\IfNoValueF{#1}{({#1},{#2},{#3},{#4},{#5})}}\xspace}
\NewDocumentCommand{\dprfContrib}{ g g }{\ensuremath{\mathsf{dprfContrib}\IfNoValueF{#1}{({#1},{#2})}}\xspace}
\NewDocumentCommand{\dprfContribution}{ g }{\ensuremath{d\IfNoValueF{#1}{_{#1}}}\xspace}
\NewDocumentCommand{\dprfVerify}{ g g g }{\ensuremath{\mathsf{dprfVerify}\IfNoValueF{#1}{({#1},{#2},{#3})}}\xspace}
\NewDocumentCommand{\dprfEval}{ g g }{\ensuremath{\mathsf{dprfEval}\IfNoValueF{#1}{({#1},{#2})}}\xspace}
\NewDocumentCommand{\dprfContribOracle}{ m m g }{\ensuremath{\mathcal{O}_{{#1},{#2}}\IfNoValueF{#3}{.{#3}}}\xspace}
\NewDocumentCommand{\dprfContribCmd}{ g }{\ensuremath{\mathit{contrib}\IfNoValueF{#1}{({#1})}}\xspace}
\NewDocumentCommand{\dprfTestOracle}{ m g }{\ensuremath{\mathcal{O}_{#1}^{?}\IfNoValueF{#2}{({#2})}}\xspace}
\NewDocumentCommand{\dprfRealOracle}{ m g }{\ensuremath{\mathcal{O}_{#1}^{\mathrm{real}}\IfNoValueF{#2}{({#2})}}\xspace}
\NewDocumentCommand{\dprfRandomOracle}{ m g }{\ensuremath{\mathcal{O}_{#1}^{\mathrm{rand}}\IfNoValueF{#2}{({#2})}}\xspace}

\newcommand{\vssScheme}{\ensuremath{\mathcal{V}}\xspace}
\newcommand{\vssModulus}{\ensuremath{q}\xspace}
\newcommand{\vssPubKey}[1]{\ensuremath{\mathit{vpk}_{#1}}\xspace}
\newcommand{\vssPrivKey}[1]{\ensuremath{\mathit{vsk}_{#1}}\xspace}
\newcommand{\vssSecParam}{\ensuremath{\kappa}\xspace}
\newcommand{\vssSecretAlt}{\ensuremath{\alt{s}}\xspace}
\newcommand{\vssShareAlt}[1]{\ensuremath{\alt{u}_{#1}}\xspace}

\newcommand{\vssCommitmentAlt}{\ensuremath{\alt{c}}\xspace}
\newcommand{\vssCommitmentAltAlt}{\ensuremath{\altalt{c}}\xspace}
\newcommand{\vssFailure}{\ensuremath{\bot}\xspace}
\newcommand{\vssHidingAdversary}[1]{\ensuremath{\mathcal{A}_{#1}}\xspace}
\newcommand{\vssBindingAdversary}[1]{\ensuremath{\mathcal{A}_{#1}}\xspace}
\newcommand{\vssNoShare}{\ensuremath{\bot}\xspace}

\NewDocumentCommand{\vssInit}{ g g g }{\ensuremath{\mathsf{vssInit}\IfNoValueF{#1}{({#1},{#2},{#3})}}\xspace}
\NewDocumentCommand{\vssShareSecret}{ g g g }{\ensuremath{\mathsf{vssShare}\IfNoValueF{#1}{({#1},{#2},{#3})}}\xspace}
\NewDocumentCommand{\vssVerify}{ g g g }{\ensuremath{\mathsf{vssVerify}\IfNoValueF{#1}{({#1},{#2},{#3})}}\xspace} 
\NewDocumentCommand{\vssReconstruct}{ g g }{\ensuremath{\mathsf{vssReconstruct}\IfNoValueF{#1}{({#1},{#2})}}\xspace}
\NewDocumentCommand{\vssCombineCommitments}{ g g }{\ensuremath{\mathsf{vssCombineCommitments}\IfNoValueF{#1}{({#1},{#2})}}\xspace}
\NewDocumentCommand{\vssCommitment}{ g }{\ensuremath{c\IfNoValueF{#1}{({#1})}}\xspace}
\NewDocumentCommand{\vssShare}{ g g }{\ensuremath{u_{{#1}\IfNoValueF{#2}{{#2}}}}\xspace}
\NewDocumentCommand{\vssMakePoly}{ g g }{\ensuremath{\mathsf{vssMakeSecret}\IfNoValueF{#1}{({#1},{#2})}}\xspace}

\NewDocumentCommand{\vssMultShares}{ g g g g g g}{\ensuremath{\mathsf{vssMultShares}\IfNoValueF{#1}{({#1},{#2},{#3},{#4},{#5},{#6})}}\xspace}
\NewDocumentCommand{\vssMultVerify}{ g g g g }{\ensuremath{\mathsf{vssMultVerify}\IfNoValueF{#1}{({#1},{#2},{#3},{#4})}}\xspace} 
\NewDocumentCommand{\vssMultReconstruct}{ g g }{\ensuremath{\mathsf{vssMultReconstruct}\IfNoValueF{#1}{({#1},{#2})}}\xspace}

\NewDocumentCommand{\vssReplicaOracle}{ m m g }{\ensuremath{\mathcal{O}_{{#1},{#2}}\IfNoValueF{#3}{.{#3}}}\xspace}
\NewDocumentCommand{\vssHidingOracle}{ m m g }{\ensuremath{\mathcal{O}_{#1}^{#2}\IfNoValueF{#3}{.{#3}}}\xspace}
\newcommand{\vssHidingBit}{\ensuremath{b}\xspace}

\newcommand{\vssSchemeNew}{\ensuremath{\mathcal{V}^*}\xspace}
\newcommand{\vssPubKeyNew}[1]{\ensuremath{\mathit{vpk}_{#1}^*}\xspace}
\newcommand{\vssPrivKeyNew}[1]{\ensuremath{\mathit{vsk}_{#1}^*}\xspace}
\newcommand{\vssRecoveryShareNew}[1]{\ensuremath{v_{#1}^*}\xspace}

\newcommand{\maskingPolyX}[1]{\ensuremath{x_{#1}}\xspace}
\newcommand{\maskingPolyY}[1]{\ensuremath{y_{#1}}\xspace}
\newcommand{\maskingPolyIdx}{\ensuremath{j}\xspace}
\newcommand{\maskingPolyPoints}[1]{\ensuremath{\mathsf{Points}_{#1}}\xspace}
\newcommand{\maskingPoly}[1]{\ensuremath{s_{#1}}\xspace}

\NewDocumentCommand{\vssInitNew}{ g g g }{\ensuremath{\mathsf{vssInit}^*\IfNoValueF{#1}{({#1},{#2},{#3})}}\xspace}
\NewDocumentCommand{\vssShareSecretNew}{ g g g g }{\ensuremath{\mathsf{vssShare}^*\IfNoValueF{#1}{({#1},{#2},{#3},{#4})}}\xspace}
\NewDocumentCommand{\vssVerifyNew}{ g g g g }{\ensuremath{\mathsf{vssVerify}^*\IfNoValueF{#1}{({#1},{#2},{#3},{#4})}}\xspace} 
\NewDocumentCommand{\vssShareNew}{ g g }{\ensuremath{u_{#1}^*\IfNoValueF{#2}{[{#2}]}}\xspace}
\NewDocumentCommand{\vssShareNewAlt}{ g g }{\ensuremath{\alt{u}_{#1}^*\IfNoValueF{#2}{[{#2}]}}\xspace}
\NewDocumentCommand{\vssCommitmentNew}{ g g }{\ensuremath{c^*\IfNoValueF{#1}{[#1]}}\xspace}
\NewDocumentCommand{\vssReconstructNew}{ g g g }{\ensuremath{\mathsf{vssReconstruct}^*\IfNoValueF{#1}{({#1},{#2},{#3})}}\xspace}
\NewDocumentCommand{\vssRecoverContribNew}{ g g g g g }{\ensuremath{\mathsf{vssRecoverContrib}^*\IfNoValueF{#1}{({#1},{#2},{#3},{#4},{#5})}}\xspace}
\NewDocumentCommand{\vssRecoverVerifyNew}{ g g g g g }{\ensuremath{\mathsf{vssRecoverVerify}^*\IfNoValueF{#1}{({#1},{#2},{#3},{#4},{#5})}}\xspace}
\NewDocumentCommand{\vssRecoverNew}{ g g g g g }{\ensuremath{\mathsf{vssRecover}^*\IfNoValueF{#1}{({#1},{#2},{#3},{#4},{#5})}}\xspace}
\NewDocumentCommand{\vssBlindedShare}{ g }{\ensuremath{u\IfNoValueF{#1}{_{#1}}}\xspace}
\NewDocumentCommand{\vssBlindedCommitment}{ g }{\ensuremath{c\IfNoValueF{#1}{_{#1}}}\xspace}
\NewDocumentCommand{\vssSecret}{ g }{\ensuremath{s\IfNoValueF{#1}{_{#1}}}\xspace}
\NewDocumentCommand{\vssContribCmd}{ g }{\ensuremath{\mathit{contrib}\IfNoValueF{#1}{({#1})}}\xspace}
\NewDocumentCommand{\vssRecoverCmd}{ g g }{\ensuremath{\mathit{recover}\IfNoValueF{#1}{({#1},{#2})}}\xspace}
\newcommand{\vssCompromiseCmd}{\ensuremath{\mathit{compromise}}\xspace}

\NewDocumentCommand{\pedVssGenerator}{ }{\ensuremath{g}\xspace}
\NewDocumentCommand{\pedVssGroupElem}{ }{\ensuremath{h}\xspace}
\NewDocumentCommand{\pedVssGroupOrder}{ }{\ensuremath{q}\xspace}
\NewDocumentCommand{\pedVssMasking}{ g }{\ensuremath{t\IfNoValueF{#1}{_{#1}}}\xspace}
\NewDocumentCommand{\pedVssMaskingAlt}{ g }{\ensuremath{\alt{t}\IfNoValueF{#1}{_{#1}}}\xspace}
\NewDocumentCommand{\pedBivPoly}{ }{\ensuremath{\hat{s}}\xspace}
\NewDocumentCommand{\eAVSSPoly}{ g }{\ensuremath{\hat{s}_{#1}}\xspace}
\NewDocumentCommand{\KatePoly}{ g }{\ensuremath{\hat{s}_{#1}}\xspace}
\NewDocumentCommand{\kateVssGroupOrder}{ }{\ensuremath{q}\xspace}

\NewDocumentCommand{\putkey}{ g g }{\ensuremath{\mathsf{put}\IfNoValueF{#1}{({#1},{#2})}}\xspace}
\NewDocumentCommand{\getkey}{ g }{\ensuremath{\mathsf{get}\IfNoValueF{#1}{({#1})}}\xspace}

\NewDocumentCommand{\residue}{ g }{\ensuremath{v\IfNoValueF{#1}{_{#1}}}\xspace}

\begin{document}

\title{sAVSS: Scalable Asynchronous Verifiable Secret Sharing in BFT Protocols}

\author[1,2,3]{Soumya Basu}
\author[1,4]{Alin Tomescu}
\author[1]{Ittai Abraham}
\author[1]{Dahlia Malkhi}
\author[1,5]{Michael K. Reiter}
\author[2,3]{Emin G{\"u}n Sirer}
\affil[1]{VMWare Research}
\affil[2]{Cornell University}
\affil[3]{IC3}
\affil[4]{MIT}
\affil[5]{UNC-Chapel Hill}

\maketitle

\begin{abstract}
This paper introduces a new way to incorporate verifiable secret sharing (VSS) schemes into Byzantine Fault Tolerance
(BFT) protocols.
This technique extends the threshold guarantee of classical Byzantine Fault Tolerant algorithms to include privacy as
well.
This provides applications with a powerful primitive: a threshold trusted third party, which simplifies many difficult
problems such as a fair exchange.

In order to incorporate VSS into BFT, we introduced \ourname{}, a
framework that transforms any VSS scheme into an asynchronous VSS
scheme with constant overhead.  By incorporating Kate et al.'s
scheme~\cite{kate2010constant-size} into our framework, we obtain an
asynchronous VSS that has constant overhead on each replica---the
first of its kind.

We show that a key-value store built using BFT replication and \ourname{}
supports writing secret-shared values with about a $30\%-50\%$
throughput overhead with less than $35$ millisecond request latencies.
\end{abstract}

\IEEEpeerreviewmaketitle

\section{Introduction}

Combining the power of Byzantine Fault Tolerant (BFT) replication with secret sharing, one can build a
decentralized service that acts over private values in a coordinated manner by consensus decrees.
This powerful combination can be leveraged in various ways to build an
automated, decentralized threshold trusted third party (T3P). For example, it may be used
to build a decentralized T3P escrow. Crucially, escrowed secrets 
will be opened by consensus decree, not necessarily requiring client interaction.
One can easily derive a fair-exchange from such
an escrow service: One party confidentially stores one value, another party
confidentially stores a second value; a consensus decree opens both. 
Another example use-case is a decentralized T3P certification authority (CA). The CA
employs some policy that automates certification decisions. The CA utilizes
threshold signing to certify documents, and again, if and when to certify is
decided automatically by consensus. 
Using polynomial secret sharing, multiple values entrusted to a decentralized T3P may be 
aggregated without client involvement. Simple additive aggregations are trivial
to implement. Arbitrary multi-party computation is possible (e.g.,~\cite{damgard2012mpc, nordholt2018minimising}), though more costly. 

In all of these use-cases, the enabling core is a mechanism called
Verifiable Secret Sharing (VSS)~\cite{chor1985verifiable} used for populating a decentralized
service with secret values.  Our use of VSS weaves it 
into BFT replication in order to automate the handling of secret values. 
For example, in a private key-value store we designed, a client request to store
an entry is broken into two parts, public and private. The public part works as
a normal BFT replication request. However, replicas delay their participation in
the ordering protocol until they obtain a verifiable share of the private
component of the store request. The private part of a client protocol 
utilizes VSS for the client to directly share the private entry.

Partly due to the need to weave VSS into a BFT replication engine, 
the setting of interest to us is asynchronous.
Relying on synchrony requires making conservative assumptions about the network delay, whereas
asynchronous protocols move at the speed of the network.
Even worse, incorporating a synchronous VSS scheme into a BFT replication engine
would require the replication engine to make a synchrony assumption even if one was not
required for the BFT protocol to work.

The best known Asynchronous VSS (AVSS) solutions 
require a client (dealer) to incur
quadratic communication and message complexities in order to store a single value~\cite{kate2010constant-size}.
This requires each replica to process and store a linear number of bits, which means that
the performance overhead due to the addition of secret sharing increases linearly with the
number of replicas.
When AVSS and BFT replication solutions were originally developed, most BFT
solutions were aimed for systems of four ($f=1$) or seven ($f=2$) replicas.
However, today,
BFT replication is being revisited at scale in blockchain systems of hundreds or
thousands of replicas.
Incurring such a large degradation in service performance for privacy may be prohibitively expensive.

To scale out AVSS, this paper introduces a new verifiable secret sharing framework called \ourname{}.
\ourname{} is a framework that, given a VSS scheme, constructs an AVSS scheme whose performance is only
a constant factor away from the original VSS scheme.
We instantiate \ourname{} in two ways: (1) using Kate, et al.'s secret sharing scheme~\cite{kate2010constant-size},
which gives us a an AVSS that has constant time share verification and share recovery and (2) using Pedersen's secret
sharing scheme~\cite{pedersen1991non-interactive} which, while only providing linear time share verification and
recovery, has cheaper cryptographic operations.
Our framework is based on one key concept: the recovery polynomial.
The recovery polynomial is a single polynomial that encodes recovery information for $\numfaulty$
shares.
Thus, by only sharing a small, constant number of additional polynomials, the client can enable all
$3\numfaulty + 1$ shares to be recovered.

We intertwine \ourname{} in a BFT replication system and build a full private key-value store solution. 
Our key-value store performs well in practice, with only a $30\%$ to $50\%$ throughput overhead over
a nonprivate key-value store with request latencies less than $35$ milliseconds.

This paper contributes a new framework for constructing asynchronous verifiable secret sharing schemes through
the use of recovery polynomials, \ourname{}.
We then instantiate \ourname{} using two verifiable secret sharing schemes and benchmark the overhead that our
new framework adds.
Finally, we incorporate our two instantiations into PBFT~\cite{castro2002practical} and evaluate a private,
Byzantine Fault Tolerant key value store.

\section{Technical Overview}
\label{sec:overview}

In this section, we provide a high-level, informal overview of the core
technique we develop to solve the asynchronous VSS problem. A precise
description and details are given in the following sections.

\subsection{The Asynchronous VSS Problem}

In the asynchronous VSS problem, a dealer shares to a group of
\numreplicas participants a polynomial \vssSecret. The API for sharing
is denoted \vssShareSecret.  If the sharing completes anywhere, then
eventually every non-faulty participant completes the sharing.

The basic method for secret sharing (API: \vssShareSecret) is to provide participant
\replicaIdx a point $(x_{\replicaIdx} , \vssSecret(x_{\replicaIdx}) )$
on the polynomial \vssSecret. The method fulfills two key properties,
\emph{hiding} and \emph{binding}:

\begin{itemize}
\item
Loosely speaking, hiding means that for a polynomial \vssSecret of
degree \numfaulty, any $\thresh = \numfaulty+1$ shares suffice to
reconstruct it via interpolation (API: \vssReconstruct), and that no
combination of \numfaulty or less reveal any information about it.

\item
Binding means that every participant receives, in addition to its
private share, a global commitment \vssCommitment to the polynomial
\vssSecret that binds the share it receives as a verifiable valid
share of \vssSecret (API: \vssVerify).

\end{itemize}

In asynchronous settings, a dealer can wait for at most $\numreplicas
- \numfaulty$ participants to acknowledge receiving a valid share,
before it inevitably may walk away.
Note that it is possible for the dealer to walk away before all of the
honest replicas have a valid share.
The asynchronous VSS problem requires that if the dealer (or any participant) 
completes the share protocol, then every correct participant can eventually 
reconstruct its share using a distributed protocol with $\numfaulty+1$ correct
participants: Participants contribute recovery information (API:
\vssRecoverContribNew), which is validated by the recipient (API:
\vssRecoverVerifyNew) and then combined to reconstruct the missing
share (API: \vssRecoverNew).

\textbf{AVSS in Byzantine Fault Tolerance}
There are a few design goals to meet when using AVSS for state machine replication.
For example, it is acceptable for a Byzantine client to lose the hiding guarantee.
However, every sharing must always be binding otherwise the replicated state machine
can be in an inconsistent state.

Additionally, there are many different Byzantine Fault Tolerance (BFT) algorithms in the literature 
that have been optimized to perform under certain circumstances.
For example, some BFT algorithms~\cite{kotla2008zyzzyva,golan-gueta2018sbft} have often incorporated a linear
"fast-path" suitable for cases where there are few failures.
In particular, this search for more optimized performance in the common case is something
that we forsee continuing in the BFT literature.

Thus, it is important for a secret sharing scheme to have minimal overheads in the common case.
In particular, a verifiable secret sharing scheme used in BFT must meet the requirement that
\vssShareSecretNew only incurs $O(1)$ overhead for the replicas.
This ensures that the same techniques will be reusable for more scalable BFT protocols that
work with larger clusters.

\subsection{Existing Solutions.}
\label{sec:overview:existing}

The seminal work by Shamir~\cite{shamir1979how} introduced the idea of employing
polynomial interpolation, a technique that was used before for error correction
codes, to share a secret with unconditional security.
A line of work emanated from this result and addressed many additional features,
such as share verifiability, asynchrony, and proactive share refresh.

Share verifiability tackles the problem of a malicious dealer that equivocates
and maliciously shares values that are inconsistent.
There are many such schemes with different properties, from classical works such
as Feldman~\cite{feldman1987practical} and Pedersen~\cite{pedersen1991non-interactive}'s
schemes to newer works such as Kate et al.~\cite{kate2010constant-size} and 
SCAPE~\cite{cascudo2017scrape}. \ourname{} can take any of these works as input and
construct a verifiable secret sharing scheme that also handles asynchrony.

Original solutions for asynchronous VSS in the information-theoretic
setting were introduced in the context of Byzantine agreement and
secure MPC~\cite{canetti1993fast}. They incur communication
complexity of $O(\numreplicas^6 \log \numreplicas)$ and message
complexity $O(\numreplicas^5)$.

% ...
\textbf{AVSS.}
The first practical asynchronous VSS solution in the computational
setting was introduced by Cachin et
al.~\cite{cachin2002asynchronous}. We will refer to it by the name
AVSS.  To cope with asynchrony, AVSS uses a bivariate secret
polynomial $\pedBivPoly(\cdot, \cdot)$.  Share \replicaIdx consists of
two univariate polynomials, $\pedBivPoly(\replicaIdx, \cdot)$,
$\pedBivPoly(\cdot, \replicaIdx)$, and so the dealer sends
$O(\numreplicas)$ information to each participant.  A missing
\replicaIdx'th share can be reconstructed from $\numfaulty+1$
evaluations of $\pedBivPoly(\replicaIdx, \cdot)$ and $\numfaulty+1$
evaluations of $\pedBivPoly(\cdot, \replicaIdx)$, incurring linear
communication overhead per recovery, for an overall recovery
complexity of $O(\numreplicas^2)$ messages and $O(\numreplicas^3)$
bits.

Additionally, participants need to verify that all shares are bound to
the same polynomial. AVSS makes use of Pedersen polynomial
commitments~\cite{pedersen1991non-interactive} for all polynomials
$\pedBivPoly(\replicaIdx, \cdot)$, $\pedBivPoly(\cdot, \replicaIdx)$,
$\replicaIdx=1..\numreplicas$.  This commitment scheme leverages the
hardness of discrete log in a multiplicative group of order
\pedVssGroupOrder with generator \pedVssGenerator.  A commitment
\vssCommitment{\residue} to a value $\residue \in
\residues{\pedVssGroupOrder}$ is a value $\pedVssGenerator^{\residue}
\pedVssGroupElem^{\nonce}$, where \pedVssGroupElem is another element
of the group and \nonce is a secret drawn at random from
\residues{\pedVssGroupOrder}.  A Pedersen commitment to a polynomial
$\vssSecret(\cdot) \in \polynomials{\pedVssGroupOrder}$ consists of a
set of commitments to \numreplicas values, i.e.,
$\vssCommitment{\vssSecret(\cdot)} = \{\langle \polyX_{\replicaIdx},
\vssCommitment{\vssSecret(\polyX_{\replicaIdx})}\rangle\}_{\replicaIdx=1}^{\numreplicas}$.
Given any pair $\langle \polyX, \vssSecret(\polyX)\rangle$, it is
possible to verify that this point is on $\vssSecret(\cdot)$ using the
commitment's homomorphic properties, i.e., that for any $\residue{1},
\residue{2} \in \residues{\pedVssGroupOrder}$,
$\vssCommitment(\residue{1})\vssCommitment(\residue{2})$ is a valid
commitment to $\residue{1} + \residue{2} \bmod \pedVssGroupOrder$.

AVSS weaves into the sharing protocol the dissemination of commitments~\cite{bracha1985asynchronous}
while incurring message complexity $O(\numreplicas^2)$ and
communication complexity $O(\numreplicas^3)$.

% ...
\textbf{eAVSS-SC.}
Kate et al.~\cite{kate2010constant-size} introduces a polynomial
commitment that has constant size. This commitment scheme leverages the
hardness of the $q$-Strong Diffe-Hellman assumption in some group with order
\kateVssGroupOrder where $\vssGen$ is a generator.
In Kate et al., a commitment $\vssCommitment{\vssSecret(\cdot)}$ is defined as 
$\vssCommitment{\vssSecret(\cdot)} = \vssGen^{\vssSecret(\bltrap)}$,
where \bltrap{} is unknown to all participants and $\vssSecret(\bltrap)$ is the polynomial
$\vssSecret{}$ evaluated at \bltrap{}.
To commit to a particular evaluation (or share) $\vssSecret(\replicaIdx)$, the dealer
also produces a witness, $\vssGen^{\frac{\vssSecret(\bltrap) -
\vssSecret(\replicaIdx)}{\bltrap - \replicaIdx}}$.
Given any triple of share, witness and commitment, it is possible to verify that the share
is indeed the evaluation of the polynomial at that point using a bilinear map.
The technique was employed by Backes et
al.~\cite{backes2013asynchronous} to construct an asynchronous VSS
scheme called eAVSS-SC that incurs both message and communication
complexities $O(\numreplicas^2)$.

In eAVSS-SC, a dealer chooses ,in addition to the secret polynomial
\vssSecret, another \numreplicas polynomials \eAVSSPoly{\replicaIdx},
$\replicaIdx=1..\numreplicas$.  \eAVSSPoly{\replicaIdx} encodes
share \replicaIdx of \vssSecret for recovery purposes.  Each of
\vssSecret, \eAVSSPoly{\replicaIdx}, has a constant-size polynomial
commitment due to the scheme by Kate et
al.~\cite{kate2010constant-size}. The commitments are constructed such that
a commitment of \eAVSSPoly{\replicaIdx} validates it as a share of \vssSecret.
Using the homomorphism of the commitments, eAVSS-SC weaves
into the sharing protocol the dissemination of commitments while
incurring both message complexity and bit complexity
$O(\numreplicas^2)$.

% ...
\textbf{The need for a new scheme.}
All of the above schemes fail to satisfy our two requirements above.
In particular, the dealer computes $O(\numreplicas^2)$
commitment values and sends $O(\numreplicas^2)$ bits. Hence, to date,
all asynchronous VSS solutions require a dealer, who wants to store a
single secret to a system, to incur $O(\numreplicas^2)$
communication complexities, which means that a replica must process
$O(\numreplicas)$ bits per request. This can be quite prohibitive for
moderate \numreplicas values, e.g., $\numreplicas=1,000$, and
infeasible for $\numreplicas=50,000$.

Fundamentally, prior asynchronous VSS schemes allow share recovery by
having the dealer enumerate all pairwise responses between replicas
during recovery.  In other words, if replica \replicaIdx is helping
replica \replicaIdxAlt recover, the dealer has shared with \replicaIdx
the response to send to \replicaIdxAlt.  However, such an approach
must require quadratic bandwidth on the dealer.  To get
around this difficulty, we make use of a distributed pseudorandom
function (DPRF), allowing these recovery responses to be generated
dynamically using information shared in the setup phase.

\subsection{\ourname}
\label{sec:overview:ours}

Our solution, named \ourname\ (for `\ournameexplained'), is the first in which the replica work is
constant per sharing in the common case.

Our approach is to use proactive secret sharing~\cite{herzberg1995proactive} in order to construct
recovery polynomials to help a replica recover a share.
Informally, suppose that a replica $\replicaIdx$ has share $\vssSecret(\replicaIdx)$, which is simply a point
on the polynomial $\vssSecret{}$.
In order to recover this share, the other replicas construct a \emph{Recovery Polynomial} (RP).
A Recovery Polynomial \maskingPoly{\replicaIdx} is random at every point except for \replicaIdx,
where $\maskingPoly{\replicaIdx}(\replicaIdx) = 0$.
Thus, if the recovering replica \replicaIdx receives shares of the sum of the original polynomial
and the masking polynomial, $\maskingPoly{\replicaIdx}(\cdot) + \vssSecret(\cdot)$, it can recover its
own share without obtaining any information about any other share.

However, adapting this technique to the asynchonous setting is nontrivial.
First, each RP must be well defined for each \replicaIdx and having every honest replica
agree on the same polynomial in the presence of adversarial nodes is very expensive.
We resolve this by having our dealer construct the recovery polynomials for each replica.
Since we are already checking that the dealer is sharing the secret consistently, doing this
for the recovery polynomials is easy.
Additionally, a dishonest dealer does not need any privacy guarantees, which means that
we only need to check for malicious behavior that hurts the consistency of the replica's shares.

While this approach solves the need to agree on a recovery polynomial, the dealer still needs to
share $\numreplicas$ recovery polynomials which will incur a quadratic cost.
To fix this problem, we first make the observation that there is no particular need for the 
constraint that $\maskingPoly{\replicaIdx}(\replicaIdx) = 0$ as long as replica \replicaIdx
knows what the value of $\maskingPoly{\replicaIdx}(\replicaIdx)$ is.
Our scheme uses a distributed pseudorandom function in order to communicate the value of that
point efficiently.
Now that the value of $\maskingPoly{\replicaIdx}(\replicaIdx)$ can be any random point,
we can actually encode \emph{multiple} replica's recovery polynomials as one recovery polynomial.
To ensure that all points of $\maskingPoly{\replicaIdx}$ are random, we encode
\numfaulty points into one recovery polynomial.
Thus, the dealer only needs to construct four recovery polynomials for the entire cluster.

We now present a high level description of our protocol.
Let a Recovery Polynomial \maskingPoly{\maskingPolyIdx} encode \numfaulty secret values. The
dealer partitions the secret shares of \vssSecret into $\badfraction =
\lceil \numreplicas/\numfaulty\rceil$ groups, and uses \badfraction
RPs \maskingPoly{\maskingPolyIdx}, $\maskingPolyIdx =
1..\badfraction$, to encode the corresponding groups. Every one of
original \numreplicas shares of \vssSecret is encoded in one of
the RPs. The dealer shares both \vssSecret and the
\maskingPoly{\maskingPolyIdx}'s among the \numreplicas participants,
and participants use the \maskingPoly{\maskingPolyIdx}'s for share recovery.

More specifically, an RP \maskingPoly{\maskingPolyIdx} is a random
polynomial of degree \numfaulty that has \numfaulty pre-defined
points.  For $(\maskingPolyIdx-1) \numfaulty \le \replicaIdx <
\maskingPolyIdx \numfaulty$, the recovery polynomial
\maskingPoly{\maskingPolyIdx} is constructed so that
$\maskingPoly{\maskingPolyIdx}(\replicaIdx) =
\maskingPolyY{\replicaIdx}$, where $\maskingPolyY{\replicaIdx} =
\dprfScheme(\replicaIdx)$ for a DPRF \dprfScheme with reconstruction
threshold \numfaulty.  (In our actual construction,
$\maskingPolyY{\replicaIdx} = \dprfScheme(\langle \nonce, \replicaIdx
\rangle)$ for a random value \nonce, to ensure that
\maskingPoly{\maskingPolyIdx} is distinct each time.  But we elide
\nonce for our discussion here.)  The dealer shares each
\maskingPoly{\maskingPolyIdx} among the \numreplicas participants, as
well as \vssSecret.

To recover its share, participant \replicaIdx probes other
participants, to which each participant responds with its share of
$\vssSecret + \maskingPoly{\maskingPolyIdx}$ for $\maskingPolyIdx =
\lceil\replicaIdx/\numfaulty\rceil$.  Each participant \replicaIdxAlt
can construct its response from its shares of \vssSecret and
\maskingPoly{\maskingPolyIdx}.  In addition, participant \replicaIdxAlt also
responds with their shares of
$\dprfScheme(\replicaIdx)$, i.e., of the secret value
$\maskingPoly{\maskingPolyIdx}(\replicaIdx)$.  (In API terms,
\vssRecoverContribNew returns a share of $\vssSecret +
\maskingPoly{\maskingPolyIdx}$ and a share of
$\dprfScheme(\replicaIdxAlt)$.)  Participant \replicaIdxAlt then
reconstructs $\vssSecret + \maskingPoly{\maskingPolyIdx}$ in full and
computes $(\vssSecret + \maskingPoly{\maskingPolyIdx})(\replicaIdx)
- \dprfScheme(\replicaIdx) = \vssSecret(\replicaIdx)$ (API:
\vssRecoverNew).

To verify a recovery share (\vssRecoverVerifyNew), first participant
\replicaIdx validates each share of $\vssSecret +
\maskingPoly{\maskingPolyIdx}$ that it receives against the commitment
\vssCommitment{\vssSecret + \maskingPoly{\maskingPolyIdx}}, which it
computes from \vssCommitment{\vssSecret} and
\vssCommitment{\maskingPoly{\maskingPolyIdx}}. Then it validates the
recovery result against the commitment \vssCommitment{\vssSecret}.
A technicality to note here is that in the Kate et al.\ commitment
scheme, each share must be accompanied by a \emph{witness} used for
commitment validation. Participants need to recover witnesses for
validation. The witness for participant for \replicaIdx could be
encoded into the RPs, at the expense of sharing polynomials over much
larger fields.  Luckily, this is not necessary; witnesses can be
reconstructed leveraging homomorphism over the witnesses of the
participants that participate in the recovery protocol (details in \secref{sec:impl:vss:kate}
).

If validation fails, then participant \replicaIdx can prove to the
other participants that the dealer is bad. In that case, different
from AVSS and eAVSS-SC, participants expose the dealer's secret.

The complexities incurred by different participants at different steps of the
\ourname\
protocol instantiated with Kate et al.~\cite{kate2010constant-size} are as follows. 
A dealer provides each of \numreplicas participants shares and
constant-size commitments on $\badfraction+1$ polynomials. The total
communication complexity is $O(\badfraction\numreplicas)$, or simply
$O(\numreplicas)$ in the usual case where $\badfraction$ is a (small) constant. Upon each
recovery request, a participant sends a constant amount of information
to the requestor, for a total $O(t)$ communication for $t$
requests. Finally, each participant requiring share recovery obtains
shares from other participants incurring $O(\numreplicas)$ communication,
for a total $O(t\numreplicas)$ communication for $t$ requests.

\section{Share Recovery in Verifiable Secret Sharing}
\label{sec:recovery}

In this section, we detail our VSS protocol and its security.  We
begin with the definitions of distributed pseudorandom functions and
verifiable secret sharing in \secref{sec:recovery:dprf} and
\secref{sec:recovery:vss}, respectively.  We will then detail our goals
(\secref{sec:recovery:goals}), further assumptions on which our scheme
builds (\secref{sec:recovery:assumptions}), our construction
(\secref{sec:recovery:construction}) and its security
(\secref{sec:recovery:security}).

Note that our proofs are applicable to \emph{any} schemes that satisfy
our descriptions below.
Our particular instantiations are described in \secref{sec:impl}.
To highlight the generality of our descriptions, we instantiate our secret 
sharing scheme described in \secref{sec:recovery:vss} in two ways~\cite{kate2010constant-size,
pedersen1991non-interactive}, of which one gives us the desired asymptotic complexity
while the other uses more inexpensive cryptographic operations.

\subsection{Distributed Pseudorandom Functions}
\label{sec:recovery:dprf}

A distributed pseudorandom function (DPRF) is a pseudorandom function
that requires the cooperation of \thresh{} replicas out of
\numreplicas{} total replicas to evaluate~\cite{naor1999distributed}.
A DPRF \dprfScheme provides the following interfaces, where
$\nats{\numreplicas} = \{1, \ldots, \numreplicas\}$.

\begin{compactitem}
\item
  \dprfInit is a randomized procedure that returns a set of pairs
  $\{\langle\dprfPubKey{\replicaIdx},
  \dprfPrivKey{\replicaIdx}\rangle\}_{\replicaIdx \in \nats{\numreplicas}}
  \gets
  \dprfInit{1^{\dprfSecParam}}{\thresh}{\numreplicas}{\dprfDomain}{\dprfRange}$.
  Each \dprfPubKey{\replicaIdx} is public key, and each
  \dprfPrivKey{\replicaIdx} is its corresponding private key.

\item \dprfContrib is a randomized procedure that returns a
  \textit{contribution} $\dprfContribution \gets
  \dprfContrib{\dprfPrivKey{\replicaIdx}}{\dprfInput}$ if
  $\dprfInput\in\dprfDomain$ and failure (\dprfFailure) otherwise.

\item
  \dprfVerify is a deterministic procedure that returns a boolean
  value.  We require that
  \dprfVerify{\dprfPubKey{\replicaIdx}}{\dprfInput}{\dprfContribution}
  returns true if \dprfContribution is output from
  \dprfContrib{\dprfPrivKey{\replicaIdx}}{\dprfInput} with nonzero
  probability, for the private key \dprfPrivKey{\replicaIdx}
  corresponding to \dprfPubKey{\replicaIdx}.

\item
  \dprfEval is a deterministic procedure that returns a value
  $\dprfOutput \gets
  \dprfEval{\dprfInput}{\{\dprfContribution{\replicaIdx}\}_{\replicaIdx\in\replicaIdxSubset}}$,
  where $\dprfOutput \in \dprfRange$, if $\dprfInput \in \dprfDomain$,
  $\cardinality{\replicaIdxSubset} \ge \thresh$ and for all
  $\replicaIdx \in \replicaIdxSubset$,
  \dprfVerify{\dprfPubKey{\replicaIdx}}{\dprfInput}{\dprfContribution{\replicaIdx}}
  returns true.  Otherwise,
  \dprfEval{\dprfInput}{\{\dprfContribution{\replicaIdx}\}_{\replicaIdx\in\replicaIdxSubset}}
  returns \dprfFailure.
\end{compactitem}

Security for a distributed pseudorandom function is defined as
follows.  An adversary \dprfAdversary{\dprfScheme} is provided inputs
$\langle
\dprfPubKey{\replicaIdx}\rangle_{\replicaIdx\in\nats{\numreplicas}}$,
\thresh, \numreplicas, \dprfDomain, and \dprfRange, where
$\{\langle\dprfPubKey{\replicaIdx},
\dprfPrivKey{\replicaIdx}\rangle\}_{\replicaIdx \in
  \nats{\numreplicas}} \gets
\dprfInit{1^{\dprfSecParam}}{\thresh}{\numreplicas}{\dprfDomain}{\dprfRange}$.
In addition, \dprfAdversary{\dprfScheme} is given oracle access to
$\numreplicas+1$ oracles.  The first \numreplicas oracles, denoted
$\langle \dprfContribOracle{\dprfScheme}{\replicaIdx}
\rangle_{\replicaIdx\in\nats{\numreplicas}}$, each supports two types
of queries.  \dprfAdversary{\dprfScheme} can invoke
\dprfContribOracle{\dprfScheme}{\replicaIdx}{\dprfContribCmd{\dprfInput}},
which returns \dprfContrib{\dprfPrivKey{\replicaIdx}}{\dprfInput}, or
it can invoke
\dprfContribOracle{\dprfScheme}{\replicaIdx}{\dprfCompromiseCmd},
which returns \dprfPrivKey{\replicaIdx}.  The last oracle provided to
\dprfAdversary{\dprfScheme} is denoted $\dprfTestOracle{\dprfScheme}:
\dprfDomain \rightarrow \dprfRange$ and is instantiated as one of two
oracles, either \dprfRealOracle{\dprfScheme} or
\dprfRandomOracle{\dprfScheme}.  Oracle \dprfRealOracle{\dprfScheme},
on input \dprfInput, selects a subset $\replicaIdxSubset \subseteq
\nats{\numreplicas}$ at random of size
$\cardinality{\replicaIdxSubset} = \thresh$, invokes
$\dprfContribution{\replicaIdx} \gets
\dprfContribOracle{\dprfScheme}{\replicaIdx}{\dprfContribCmd{\dprfInput}}$
for each $\replicaIdx \in \replicaIdxSubset$, and returns
\dprfEval{\dprfInput}{\{\dprfContribution{\replicaIdx}\}_{\replicaIdx\in\replicaIdxSubset}}.
Oracle \dprfRandomOracle{\dprfScheme} is instantiated as a function
chosen uniformly at random from the set of all functions from
\dprfDomain to \dprfRange.  For any $\dprfInput \in \dprfDomain$, let
\replicaIdxSubset{\dprfInput} be the oracle indices such that for each
$\replicaIdx \in \replicaIdxSubset{\dprfInput}$,
\dprfAdversary{\dprfScheme} invokes
\dprfContribOracle{\dprfScheme}{\replicaIdx}{\dprfCompromiseCmd} or
\dprfContribOracle{\dprfScheme}{\replicaIdx}{\dprfContribCmd{\dprfInput}}.
Then, \dprfAdversary{\dprfScheme} is \textit{legitimate} if
$\cardinality{\replicaIdxSubset{\dprfInput}} < \thresh$ for every
\dprfInput for which \dprfAdversary{\dprfScheme} invokes
\dprfTestOracle{\dprfScheme}{\dprfInput}.  Finally,
\dprfAdversary{\dprfScheme} outputs a bit.  We say that the
distributed pseudorandom function is secure if for all legitimate
adversaries \dprfAdversary{\dprfScheme} that run in time polynomial in
\dprfSecParam,
\begin{equation}
\begin{array}{l@{\hspace{0.2em}}l}
& \prob{\dprfAdversary{\dprfScheme}^{\langle \dprfContribOracle{\dprfScheme}{\replicaIdx}
\rangle_{\replicaIdx\in\nats{\numreplicas}},\dprfRealOracle{\dprfScheme}}(\langle
\dprfPubKey{\replicaIdx}\rangle_{\replicaIdx\in\nats{\numreplicas}},
\thresh, \numreplicas, \dprfDomain, \dprfRange) = 1} \\
-
& \prob{\dprfAdversary{\dprfScheme}^{\langle \dprfContribOracle{\dprfScheme}{\replicaIdx}
\rangle_{\replicaIdx\in\nats{\numreplicas}},\dprfRandomOracle{\dprfScheme}}(\langle
\dprfPubKey{\replicaIdx}\rangle_{\replicaIdx\in\nats{\numreplicas}},
\thresh, \numreplicas, \dprfDomain, \dprfRange) = 1} 
\end{array}
\label{eqn:dprf}
\end{equation}
is negligible in \dprfSecParam.

\subsection{Verifiable Secret Sharing}
\label{sec:recovery:vss}

Verifiable Secret Sharing (VSS) is a way to share a secret so that it
requires a coalition of \thresh{} replicas out of \numreplicas{} total
replicas in order to reconstruct the secret.  A VSS scheme provides
the following interfaces:

\begin{itemize}
\item \vssInit is a randomized procedure that returns $\langle
  \vssModulus, \{\langle\vssPubKey{\replicaIdx},
  \vssPrivKey{\replicaIdx}\rangle\}_{\replicaIdx \in
    \nats{\numreplicas}}\rangle \gets
  \vssInit{1^{\vssSecParam}}{\thresh}{\numreplicas}$.  Here, \vssModulus
  is a prime of length \vssSecParam bits.  Each
  \vssPubKey{\replicaIdx} is a public key, and each
  \vssPrivKey{\replicaIdx} is its corresponding private key.

\item \vssShareSecret is a randomized procedure that produces
  $\langle \vssCommitment, \{\vssShare{\replicaIdx}\}_{\replicaIdx \in
    \nats{\numreplicas}}\rangle \gets
  \vssShareSecret{\vssSecret}{\vssModulus}{\{\vssPubKey{\replicaIdx}\}_{\replicaIdx\in\nats{\numreplicas}}}$.
  Here, $\vssSecret \in \polynomials{\vssModulus}$ is a degree
  $\thresh-1$ polynomial, and \vssModulus and
  $\{\vssPubKey{\replicaIdx}\}_{\replicaIdx\in\nats{\numreplicas}}$
  are as output by \vssInit.  The value \vssCommitment is a
  \textit{commitment}, and each \vssShare{\replicaIdx} is a
  \textit{share}.

\item \vssVerify is a deterministic procedure that returns a boolean.
  We require that
  \vssVerify{\vssPubKey{\replicaIdx}}{\vssCommitment}{\vssShare{\replicaIdx}}
  return true if $\langle \vssCommitment,
  \vssShare{\replicaIdx}\rangle$ (i.e., with arbitrary
  $\{\vssShare{\replicaIdxAlt}\}_{\replicaIdxAlt \neq \replicaIdx}$)
  is output from
  \vssShareSecret{\vssSecret}{\vssModulus}{\{\vssPubKey{\replicaIdx}\}_{\replicaIdx\in\nats{\numreplicas}}}
  with nonzero probability.

\item \vssReconstruct is a deterministic procedure that returns a
  value $\vssSecret \gets \vssReconstruct{\vssCommitment}{\{\langle
    \vssPubKey{\replicaIdx},
    \vssShare{\replicaIdx}\rangle\}_{\replicaIdx\in\replicaIdxSubset}}$
  where $\vssSecret \in \polynomials{\vssModulus}$ of degree
  $\thresh-1$, if $\cardinality{\replicaIdxSubset} \ge \thresh$ and
  for all $\replicaIdx \in \replicaIdxSubset$,
  \vssVerify{\vssPubKey{\replicaIdx}}{\vssCommitment}{\vssShare{\replicaIdx}}
  returns true.  Otherwise, \vssReconstruct{\vssCommitment}{\{\langle
    \vssPubKey{\replicaIdx},
    \vssShare{\replicaIdx}\rangle\}_{\replicaIdx\in\replicaIdxSubset}}
  returns \vssFailure.
\end{itemize}

The security of a VSS scheme lies in its \textit{hiding} and
\textit{binding} properties.

\subsubsection{Hiding}
A hiding adversary \vssHidingAdversary{\vssScheme} is provided inputs
\vssModulus and $\{\vssPubKey{\replicaIdx}\}_{\replicaIdx \in
  \nats{\numreplicas}}$, where $\langle \vssModulus,
\{\langle\vssPubKey{\replicaIdx},
\vssPrivKey{\replicaIdx}\rangle\}_{\replicaIdx \in
  \nats{\numreplicas}}\rangle \gets
\vssInit{1^{\vssSecParam}}{\thresh}{\numreplicas}$, and access to
$\numreplicas+1$ oracles.  The first \numreplicas oracles are denoted
$\langle \vssReplicaOracle{\vssScheme}{\replicaIdx}
\rangle_{\replicaIdx \in \nats{\numreplicas}}$; each
\vssReplicaOracle{\vssScheme}{\replicaIdx} is initialized with
\vssPrivKey{\replicaIdx} and can be invoked as described below.  The
last oracle provided to \vssHidingAdversary{\vssScheme} is denoted
\vssHidingOracle{\vssScheme}{\vssHidingBit}, where $\vssHidingBit \in
\{0,1\}$.  \vssHidingAdversary{\vssScheme} can invoke this oracle with
two inputs $\vssSecret{0}, \vssSecret{1} \in \residues{\vssModulus}$.
When invoked, \vssHidingOracle{\vssScheme}{\vssHidingBit} generates a
random $\vssSecretAlt \in \polynomials{\vssModulus}$ of degree $\thresh-1$
such that $\vssSecretAlt(0) = \vssSecret{\vssHidingBit}$ and performs
$\langle \vssCommitment,
\{\vssShare{\replicaIdx}\}_{\replicaIdx\in\nats{\numreplicas}}
\rangle \gets
\vssShareSecret{\vssSecretAlt}{\vssModulus}{\{\vssPubKey{\replicaIdx}\}_{\replicaIdx\in\nats{\numreplicas}}}$,
providing \vssCommitment to \vssHidingAdversary{\vssScheme} and
$\langle\vssCommitment, \vssShare{\replicaIdx}\rangle$ to
\vssReplicaOracle{\vssScheme}{\replicaIdx}.  The oracles
$\langle\vssReplicaOracle{\vssScheme}{\replicaIdx}\rangle_{\replicaIdx\in\nats{\numreplicas}}$
can be invoked by \vssHidingAdversary{\vssScheme} as follows.
\vssHidingAdversary{\vssScheme} can invoke
\vssReplicaOracle{\vssScheme}{\replicaIdx}{\vssContribCmd{\vssCommitment}},
which returns the share \vssShare{\replicaIdx}
provided to \vssReplicaOracle{\vssScheme}{\replicaIdx} with commitment
\vssCommitment by \vssHidingOracle{\vssScheme}{\vssHidingBit}.
\vssHidingAdversary{\vssScheme} can also invoke
\vssReplicaOracle{\vssScheme}{\replicaIdx}{\vssCompromiseCmd}, which
returns \vssPrivKey{\replicaIdx} and all $\langle \vssCommitment,
\vssShare{\replicaIdx}\rangle$ pairs
received from \vssHidingOracle{\vssScheme}{\vssHidingBit}.
For any \vssCommitment, let \replicaIdxSubset{\vssCommitment} be the
oracle indices such that for each $\replicaIdx \in
\replicaIdxSubset{\vssCommitment}$, \vssHidingAdversary{\vssScheme}
invokes \vssReplicaOracle{\vssScheme}{\replicaIdx}{\vssCompromiseCmd}
or
\vssReplicaOracle{\vssScheme}{\replicaIdx}{\vssContribCmd{\vssCommitment}}.
Then, \vssHidingAdversary{\vssScheme} is \textit{legitimate} if
$\cardinality{\replicaIdxSubset{\vssCommitment}} < \thresh$ for every
\vssCommitment.  Finally, \vssHidingAdversary{\vssScheme} outputs a
bit.  We say that the VSS \vssScheme is \textit{hiding} if for all
legitimate adversaries \vssHidingAdversary{\vssScheme} that run in
time polynomial in \vssSecParam,
\begin{equation}
\begin{array}{ll}
& \prob{\vssHidingAdversary{\vssScheme}^{\langle
      \vssReplicaOracle{\vssScheme}{\replicaIdx}
      \rangle_{\replicaIdx\in\nats{\numreplicas}},\vssHidingOracle{\vssScheme}{1}}(\vssModulus,
    \{\vssPubKey{\replicaIdx}\}_{\replicaIdx \in \nats{\numreplicas}})
    = 1} \\
  - & \prob{\vssHidingAdversary{\vssScheme}^{\langle
      \vssReplicaOracle{\vssScheme}{\replicaIdx}
      \rangle_{\replicaIdx\in\nats{\numreplicas}},\vssHidingOracle{\vssScheme}{0}}(\vssModulus,
    \{\vssPubKey{\replicaIdx}\}_{\replicaIdx \in \nats{\numreplicas}})
    = 1}
\end{array}
\label{eqn:hiding}
\end{equation}
is negligible in \vssSecParam.

\subsubsection{Binding}
A binding adversary \vssBindingAdversary{\vssScheme} is provided inputs
$\langle \vssModulus, \{\langle\vssPubKey{\replicaIdx},
\vssPrivKey{\replicaIdx}\rangle\}_{\replicaIdx \in
  \nats{\numreplicas}}\rangle \gets
\vssInit{1^{\vssSecParam}}{\thresh}{\numreplicas}$.
\vssBindingAdversary{\vssScheme} outputs \vssCommitment, $\{
\vssShare{\replicaIdx}\}_{\replicaIdx
  \in \replicaIdxSubset}$ and $\{ \vssShareAlt{\replicaIdx}\}_{\replicaIdx \in
  \replicaIdxSubsetAlt}$.  We say that VSS \vssScheme is
\textit{binding} if for all binding adversaries
\vssBindingAdversary{\vssScheme} that run in time polynomial in
\vssSecParam,
\[
\prob{\begin{array}{ll}
    & \vssReconstruct{\vssCommitment}{\{\langle \vssPubKey{\replicaIdx}, \vssShare{\replicaIdx}\rangle\}_{\replicaIdx\in\replicaIdxSubset}} = \vssSecret \\
    \wedge & \vssReconstruct{\vssCommitment}{\{\langle \vssPubKey{\replicaIdx}, \vssShareAlt{\replicaIdx}\rangle\}_{\replicaIdx\in\replicaIdxSubsetAlt}} = \vssSecretAlt \\
    \wedge & \vssSecret \neq \vssFailure \wedge \vssSecretAlt \neq \vssFailure \wedge \vssSecret \neq \vssSecretAlt
    \end{array}}
\]
is negligible in \vssSecParam, where the probability is taken with
respect to random choices made in \vssInit and by
\vssBindingAdversary{\vssScheme}.

\subsection{Goals}
\label{sec:recovery:goals}

Given such a VSS scheme \vssScheme and a DPRF \dprfScheme, our goal is
to construct a new VSS scheme \vssSchemeNew that provides the
\vssInit, \vssShareSecret, \vssVerify, and \vssReconstruct algorithms
(denoted \vssInitNew, \vssShareSecretNew, \vssVerifyNew and
\vssReconstructNew for \vssSchemeNew, respectively) as defined
in \secref{sec:recovery:dprf}, as well as three more algorithms, denoted
\vssRecoverContribNew, \vssRecoverVerifyNew, and \vssRecoverNew.  We
allow the \vssShareSecretNew algorithm to accept additional arguments
(a set of private keys for a DPRF) and to return an additional value
\nonce that is provided as input to all procedures except for
\vssInitNew.  The algorithms \vssRecoverContribNew,
\vssRecoverVerifyNew, and \vssRecoverNew together permit a replica to
recover its share from other replicas, and behave as follows:
\begin{compactitem}
\item \vssRecoverContribNew is a randomized procedure that returns
  $\vssRecoveryShareNew{\replicaIdx} \gets
  \vssRecoverContribNew{\vssCommitmentNew}{\nonce}{\vssPrivKeyNew{\replicaIdx}}{\vssShareNew{\replicaIdx}}{\replicaIdxAlt}$
  where \vssRecoveryShareNew{\replicaIdx} is a \textit{recovery share}
  with properties described below.
\item \vssRecoverVerifyNew is a deterministic procedure that returns
  a boolean.  \vssRecoverVerifyNew{\vssCommitmentNew}{\nonce}{\vssRecoveryShareNew{\replicaIdx}}{\vssPubKeyNew{\replicaIdx}}{\replicaIdxAlt} must return
  true if \vssRecoveryShareNew{\replicaIdx} is output from 
  \vssRecoverContribNew{\vssCommitmentNew}{\nonce}{\vssPrivKeyNew{\replicaIdx}}{\vssShareNew{\replicaIdx}}{\replicaIdxAlt} with nonzero probability and
  \vssVerifyNew{\vssPubKeyNew{\replicaIdx}}{\vssCommitmentNew}{\nonce}{\vssShareNew{\replicaIdx}} returns true.
\item \vssRecoverNew is a deterministic procedure that returns
  $\vssShareNew{\replicaIdxAlt} \gets
  \vssRecoverNew{\vssCommitmentNew}{\nonce}{\{\langle
    \vssPubKeyNew{\replicaIdx},
    \vssRecoveryShareNew{\replicaIdx}\rangle\}_{\replicaIdx\in\replicaIdxSubset}}{\replicaIdxAlt}{\vssPubKeyNew{\replicaIdxAlt}}$
  if $\cardinality{\replicaIdxSubset} \ge \thresh$,
  \vssRecoverVerifyNew{\vssCommitmentNew}{\nonce}{\vssRecoveryShareNew{\replicaIdx}}{\vssPubKeyNew{\replicaIdx}}{\replicaIdxAlt}
  returns true for all $\replicaIdx \in \replicaIdxSubset$, and
  \vssVerifyNew{\vssPubKeyNew{\replicaIdxAlt}}{\vssCommitmentNew}{\nonce}{\vssShareNew{\replicaIdxAlt}}
  returns true.  Otherwise,
  \vssRecoverNew{\vssCommitmentNew}{\nonce}{\{\langle
    \vssPubKeyNew{\replicaIdx},
    \vssRecoveryShareNew{\replicaIdx}\rangle\}_{\replicaIdx\in\replicaIdxSubset}}{\replicaIdxAlt}{\vssPubKeyNew{\replicaIdxAlt}}
  returns \vssFailure.
\end{compactitem}

Due to the additional interfaces above, we change the definition of
hiding security as follows.  Each oracle
\vssReplicaOracle{\vssSchemeNew}{\replicaIdx} additionally supports a
query
\vssReplicaOracle{\vssSchemeNew}{\replicaIdx}{\vssRecoverCmd{\vssCommitmentNew}{\replicaIdxAlt}}
that returns $\vssRecoveryShareNew{\replicaIdx} \gets
\vssRecoverContribNew{\vssCommitmentNew}{\nonce}{\vssPrivKeyNew{\replicaIdx}}{\vssShareNew{\replicaIdx}}{\replicaIdxAlt}$.
For any \vssCommitmentNew, let \replicaIdxSubset{\vssCommitmentNew} be
the oracle indices such that for each $\replicaIdx \in
\replicaIdxSubset{\vssCommitmentNew}$,
\vssHidingAdversary{\vssSchemeNew} invokes
\vssReplicaOracle{\vssSchemeNew}{\replicaIdx}{\vssCompromiseCmd},
\vssReplicaOracle{\vssSchemeNew}{\replicaIdx}{\vssContribCmd{\vssCommitmentNew}},
or
$\{\vssReplicaOracle{\vssSchemeNew}{\replicaIdxAlt}{\vssRecoverCmd{\vssCommitmentNew}{\replicaIdx}}\}_{\replicaIdxAlt
  \in \replicaIdxSubsetAlt}$ where $\cardinality{\replicaIdxSubsetAlt}
\ge \thresh$.  Then, \vssHidingAdversary{\vssSchemeNew} is
\textit{legitimate} if
$\cardinality{\replicaIdxSubset{\vssCommitmentNew}} < \thresh$ for
every \vssCommitmentNew.

\subsection{Assumptions on Underlying VSS}
\label{sec:recovery:assumptions}

Our construction combines an existing VSS scheme with a DPRF for
which, if $\langle \vssModulus, \{\langle\vssPubKey{\replicaIdx},
\vssPrivKey{\replicaIdx}\rangle\}_{\replicaIdx \in
  \nats{\numreplicas}}\rangle \gets
\vssInit{1^{\vssSecParam}}{\thresh}{\numreplicas}$, then $\dprfRange =
\residues{\vssModulus}$ and each share \vssShare{\replicaIdx} output
from \vssShareSecret is in \residues{\vssModulus}.  In addition, we
require that the VSS offer additional procedures, as follows.
\begin{itemize}
\item There is a procedure \vssMakePoly that creates
  \[
  \vssSecret \gets
  \vssMakePoly{\vssModulus}{\{\langle \maskingPolyX{\replicaIdx},
    \maskingPolyY{\replicaIdx} \rangle\}_{\replicaIdx \in
      \replicaIdxSubset}}
  \] where $\vssSecret \in
  \polynomials{\vssModulus}$ is of degree
  \cardinality{\replicaIdxSubset}, and so that if
  \[
  \langle \vssCommitment, \{\vssShare{\replicaIdx}\}_{\replicaIdx \in
    \nats{\numreplicas}}\rangle
    \gets \vssShareSecret{\vssSecret}{\vssModulus}{\{\vssPubKey{\replicaIdx}\}_{\replicaIdx\in\nats{\numreplicas}}}
  \]
  then $\vssShare{\replicaIdx} = \maskingPolyY{\replicaIdx}$ for
  any $\replicaIdx \in \replicaIdxSubset$.

\item There is a procedure \vssCombineCommitments
  such that if
\begin{align*}
  \vssReconstruct{\vssCommitment}{\{\langle
  \vssPubKey{\replicaIdx}, \vssShare{\replicaIdx}\rangle\}_{\replicaIdx\in\replicaIdxSubset}}
& = \vssSecret \\
\vssReconstruct{\vssCommitmentAlt}{\{\langle
    \vssPubKey{\replicaIdx}, \vssShareAlt{\replicaIdx}\rangle\}_{\replicaIdx\in\replicaIdxSubset}}
& = \vssSecretAlt
\end{align*}
where $\vssSecret, \vssSecretAlt \neq \vssFailure$, and
if 
\begin{align*}
\vssCommitmentAltAlt & \gets \vssCombineCommitments{\vssCommitment}{\vssCommitmentAlt}
\end{align*}
then
\[
\vssReconstruct{\vssCommitmentAltAlt}{\{\langle
  \vssPubKey{\replicaIdx},
  (\vssShare{\replicaIdx} + \vssShareAlt{\replicaIdx})
  \rangle\}_{\replicaIdx\in\replicaIdxSubset}}
= \vssSecret + \vssSecretAlt
\]
\end{itemize}

An example of such a scheme is that due to Kate et al.~\cite{kate2010constant-size}.

\subsection{VSS Scheme with Recovery}
\label{sec:recovery:construction}

Below we describe the procedures that make up the VSS scheme
\vssSchemeNew.  The algorithms are expressed in terms of constants
\numreplicas (the number of replicas), \thresh (the reconstruction
threshold), and $\badfraction = \lceil \numreplicas/(\thresh-1)\rceil$.
Each share \vssShareNew{\replicaIdx}
and commitment \vssCommitmentNew is a zero-indexed vector of
$\badfraction+1$ elements.  We denote the \maskingPolyIdx-th element
of each by \vssShareNew{\replicaIdx}{\maskingPolyIdx} and
\vssCommitmentNew{\maskingPolyIdx}, respectively, for $0 \le
\maskingPolyIdx \le \badfraction$.  Line numbers below refer to
\figref{fig:pcode}.

\vssInitNew initializes the underlying VSS \vssScheme in
line~\ref{line:vssInitNew:vssInit}, as well as a DPRF \dprfScheme in
line~\ref{line:vssInitNew:dprfInit}.  The public key
\vssPubKeyNew{\replicaIdx} for replica \replicaIdx consists of its
public key \vssPubKey{\replicaIdx} for \vssScheme and its public key
\dprfPubKey{\replicaIdx} for \dprfScheme
(line~\ref{line:vssInitNew:vssPubKeyNew}) and similarly for the
private key \vssPrivKeyNew{\replicaIdx}
(line~\ref{line:vssInitNew:vssPrivKeyNew}).

\begin{figure*}
  \begin{minipage}[t]{0.49\textwidth}
\begin{algorithmic}[1]
  \Procedure{\vssInitNew{1^{\vssSecParam}}{\thresh}{\numreplicas}}{}

  \State $\langle \vssModulus, \{\langle\vssPubKey{\replicaIdx},
  \vssPrivKey{\replicaIdx}\rangle\}_{\replicaIdx \in
    \nats{\numreplicas}}\rangle
  \gets \vssInit{1^{\vssSecParam}}{\thresh}{\numreplicas}$
  \label{line:vssInitNew:vssInit}
  
  \State $\langle \{\langle\dprfPubKey{\replicaIdx},
  \dprfPrivKey{\replicaIdx}\rangle\}_{\replicaIdx \in \nats{\numreplicas}} \rangle
  \gets
  \dprfInit{1^{\dprfSecParam}}{\thresh}{\numreplicas}{\{0,1\}^{\vssSecParam}\times\nats{\numreplicas}}{\residues{\vssModulus}}$
  \label{line:vssInitNew:dprfInit}

  \For{$\replicaIdx \in \nats{\numreplicas}$}
     \State $\vssPubKeyNew{\replicaIdx} \gets \langle \vssPubKey{\replicaIdx}, \dprfPubKey{\replicaIdx}\rangle$
     \State $\vssPrivKeyNew{\replicaIdx} \gets \langle \vssPrivKey{\replicaIdx}, \dprfPrivKey{\replicaIdx}\rangle$
     \label{line:vssInitNew:vssPrivKeyNew}
  \EndFor

  \State \Return $\langle \vssModulus, \{\langle\vssPubKeyNew{\replicaIdx},
  \vssPrivKeyNew{\replicaIdx}\rangle\}_{\replicaIdx\in\nats{\numreplicas}}\rangle$
  \label{line:vssInitNew:vssPubKeyNew}
  \EndProcedure

\bigskip
  
  \Procedure{\vssShareSecretNew{\vssSecret}{\vssModulus}{\{\dprfPrivKey{\replicaIdx}\}_{\replicaIdx\in\nats{\numreplicas}}}{\{\vssPubKeyNew{\replicaIdx}\}_{\replicaIdx\in\nats{\numreplicas}}}}{}
  
  \State $\nonce \getsr \{0,1\}^{\vssSecParam}$
  \label{line:vssShareSecretNew:nonce}
  \For{$\replicaIdx \in \nats{\numreplicas}$}
  \label{line:vssShareSecretNew:loop-dprfEval}
  \State $\langle \vssPubKey{\replicaIdx}, \dprfPubKey{\replicaIdx}\rangle \gets \vssPubKeyNew{\replicaIdx}$
  \State $\maskingPolyY{\replicaIdx} \gets
  \dprfEval{\langle\nonce, \replicaIdx\rangle}{\{\dprfContrib{\dprfPrivKey{\replicaIdxAlt}}{\langle\nonce, \replicaIdx\rangle}\}_{\replicaIdxAlt \in \nats{\numreplicas}}}$
  \label{line:vssShareSecretNew:dprfEval}
  \EndFor

  \For{$\maskingPolyIdx \in \nats{\badfraction}$}
  \State $\maskingPolyPoints{\maskingPolyIdx} \gets \{\langle\replicaIdx, \maskingPolyY{\replicaIdx}\rangle \mid (\maskingPolyIdx-1)(\thresh-1) < \replicaIdx \le \maskingPolyIdx(\thresh-1)\}$
  \label{line:vssShareSecretNew:maskingPolyPoints}
  \State $\maskingPoly{\maskingPolyIdx} \gets \vssMakePoly{\vssModulus}{\maskingPolyPoints{\maskingPolyIdx}}$
  \label{line:vssShareSecretNew:vssMakePoly}
  \State $\langle \vssCommitmentNew{\maskingPolyIdx}, \{\vssShareNew{\replicaIdx}{\maskingPolyIdx}\}_{\replicaIdx \in
    \nats{\numreplicas}}\rangle
    \gets \vssShareSecret{\maskingPoly{\maskingPolyIdx}}{\vssModulus}{\{\vssPubKey{\replicaIdx}\}_{\replicaIdx\in\nats{\numreplicas}}}$
    \label{line:vssShareSecretNew:vssShareDPRFVal}
  \EndFor

  \State $\langle \vssCommitmentNew{0}, \{\vssShareNew{\replicaIdx}{0}\}_{\replicaIdx \in
    \nats{\numreplicas}}\rangle
    \gets \vssShareSecret{\vssSecret}{\vssModulus}{\{\vssPubKey{\replicaIdx}\}_{\replicaIdx\in\nats{\numreplicas}}}$
    \label{line:vssShareSecretNew:vssShareVSSVal}
   
  \State \Return $\langle \vssCommitmentNew, \nonce, \{\vssShareNew{\replicaIdx}\}_{\replicaIdx \in
    \nats{\numreplicas}}\rangle$
  \label{line:vssShareSecretNew:return}
  \EndProcedure

\bigskip
  
  \Procedure{\vssVerifyNew{\vssPubKeyNew{\replicaIdx}}{\vssCommitmentNew}{\nonce}{\vssShareNew{\replicaIdx}}}{}
  \State $\langle \vssPubKey{\replicaIdx}, \dprfPubKey{\replicaIdx}\rangle \gets \vssPubKeyNew{\replicaIdx}$
  \If{$\vssVerify{\vssPubKey{\replicaIdx}}{\vssCommitmentNew{0}}{\vssShareNew{\replicaIdx}{0}} = \textrm{false}$}
  \label{line:vssVerifyNew:vssVerifyVSSShare}
  \State \Return false
  \EndIf
  \If{$\vssShareNew{\replicaIdx}{1} \neq \vssNoShare$}
  \label{line:vssVerifyNew:testRecovered}
  \State $\maskingPolyIdx \gets \lceil\replicaIdx/(\thresh-1)\rceil$
  \If{$\dprfVerify{\dprfPubKey{\replicaIdx}}{\langle\nonce, \replicaIdx\rangle}{\vssShareNew{\replicaIdx}{\maskingPolyIdx}} = \textrm{false}$}
  \label{line:vssVerifyNew:dprfVerify}
  \State \Return false
  \EndIf
  \For{$\maskingPolyIdx \in \nats{\badfraction}$}
  \If{$\vssVerify{\vssPubKey{\replicaIdx}}{\vssCommitmentNew{\maskingPolyIdx}}{\vssShareNew{\replicaIdx}{\maskingPolyIdx}} = \textrm{false}$}
  \label{line:vssVerifyNew:vssVerifyMaskingPolyShare}
  \State \Return false
  \EndIf
  \EndFor
  \EndIf
  \State \Return true
  \EndProcedure

\algstore{pcode}
\end{algorithmic}
\end{minipage}
\qquad
\begin{minipage}[t]{0.49\textwidth}
\begin{algorithmic}[1]
\algrestore{pcode}
  \Procedure{\vssReconstructNew{\vssCommitmentNew}{\nonce}{\{\langle
      \vssPubKeyNew{\replicaIdx}, \vssShareNew{\replicaIdx}
      \rangle\}_{\replicaIdx\in\replicaIdxSubset}}}{}
  \For{$\replicaIdx \in \replicaIdxSubset$}
  \If{$\vssVerifyNew{\vssPubKeyNew{\replicaIdx}}{\vssCommitmentNew}{\nonce}{\vssShareNew{\replicaIdx}} = \textrm{false}$}
  \label{line:vssReconstructNew:vssVerifyNew}
  \State \Return \vssFailure
  \EndIf
  \State $\langle \vssPubKey{\replicaIdx}, \dprfPubKey{\replicaIdx}\rangle \gets \vssPubKeyNew{\replicaIdx}$
  \EndFor
  \State \Return \vssReconstruct{\vssCommitmentNew{0}}{\{\langle
    \vssPubKey{\replicaIdx}, \vssShareNew{\replicaIdx}{0}\rangle\}_{\replicaIdx\in\replicaIdxSubset}}
  \label{line:vssReconstructNew:vssReconstruct}
  \EndProcedure

\medskip
  
  \Procedure{\vssRecoverContribNew{\vssCommitmentNew}{\nonce}{\vssPrivKeyNew{\replicaIdx}}{\vssShareNew{\replicaIdx}}{\replicaIdxAlt}}{}
  \State $\langle \vssPrivKey{\replicaIdx}, \dprfPrivKey{\replicaIdx}\rangle \gets \vssPrivKeyNew{\replicaIdx}$
  \State $\dprfContribution{\replicaIdx} \gets
  \dprfContrib{\dprfPrivKey{\replicaIdx}}{\langle\nonce,\replicaIdxAlt\rangle}$
  \label{line:vssRecoverContribNew:dprfContrib}
  \State $\maskingPolyIdx \gets \lceil\replicaIdxAlt/(\thresh-1)\rceil$
  \label{line:vssRecoverContribNew:maskingPolyIdx}
  \State \Return $\langle \dprfContribution{\replicaIdx}, (\vssShareNew{\replicaIdx}{0} + \vssShareNew{\replicaIdx}{\maskingPolyIdx}) \rangle$
  \label{line:vssRecoverContribNew:return}
  \EndProcedure

\medskip
  
  \Procedure{\vssRecoverVerifyNew{\vssCommitmentNew}{\nonce}{\vssRecoveryShareNew{\replicaIdx}}{\vssPubKeyNew{\replicaIdx}}{\replicaIdxAlt}}{}
  \State $\maskingPolyIdx \gets \lceil\replicaIdxAlt/(\thresh-1)\rceil$
  \State $\langle \dprfContribution{\replicaIdx}, \vssBlindedShare \rangle \gets \vssRecoveryShareNew{\replicaIdx}$
  \label{line:vssRecoverVerifyNew:parseVssRecoveryShareNew}
  \State $\langle \vssPubKey{\replicaIdx}, \dprfPubKey{\replicaIdx}\rangle \gets \vssPubKeyNew{\replicaIdx}$
  \If{$\dprfVerify{\dprfPubKey{\replicaIdx}}{\langle\nonce,\replicaIdxAlt\rangle}{\dprfContribution{\replicaIdx}} = \textrm{false}$}
  \label{line:vssRecoverVerifyNew:dprfVerify}
     \State \Return false
  \EndIf
  \State $\vssBlindedCommitment \gets \vssCombineCommitments{\vssCommitmentNew{0}}{\vssCommitmentNew{\maskingPolyIdx}}$
  \label{line:vssRecoverVerifyNew:vssCombineCommitments}
  \If{$\vssVerify{\vssPubKey{\replicaIdx}}{\vssBlindedCommitment}{\vssBlindedShare} = \textrm{false}$}
  \label{line:vssRecoverVerifyNew:vssVerify}
  \State \Return false
  \EndIf
  \State \Return true
  \EndProcedure

\medskip
  
  \Procedure{\vssRecoverNew{\vssCommitmentNew}{\nonce}{\{\langle
    \vssPubKeyNew{\replicaIdx},
    \vssRecoveryShareNew{\replicaIdx}\rangle\}_{\replicaIdx\in\replicaIdxSubset}}{\replicaIdxAlt}{\vssPubKeyNew{\replicaIdxAlt}}}{}
  \State $\maskingPolyIdx \gets \lceil\replicaIdxAlt/(\thresh-1)\rceil$
  \For{$\replicaIdx \in \replicaIdxSubset$}
  \If{$\vssRecoverVerifyNew{\vssCommitmentNew}{\nonce}{\vssRecoveryShareNew{\replicaIdx}}{\vssPubKeyNew{\replicaIdx}}{\replicaIdxAlt} = \textrm{false}$}
  \label{line:vssRecoverNew:vssRecoverVerifyNew}
  \State \Return \vssFailure
  \EndIf
  \State $\langle \dprfContribution{\replicaIdx}, \vssBlindedShare{\replicaIdx} \rangle \gets \vssRecoveryShareNew{\replicaIdx}$
  \State $\langle \vssPubKey{\replicaIdx}, \dprfPubKey{\replicaIdx}\rangle \gets \vssPubKeyNew{\replicaIdx}$
  \EndFor
  \State $\vssBlindedCommitment \gets \vssCombineCommitments{\vssCommitmentNew{0}}{\vssCommitmentNew{\maskingPolyIdx}}$
  \label{line:vssRecoverNew:vssCombineCommitments}
  \State $\vssSecret \gets \vssReconstruct{\vssBlindedCommitment}{\{\langle
  \vssPubKey{\replicaIdx}, \vssShare{\replicaIdx}\rangle\}_{\replicaIdx\in\replicaIdxSubset}}$
  \label{line:vssRecoverNew:vssReconstruct}

  \State $\maskingPolyY{\replicaIdxAlt} \gets \dprfEval{\langle\nonce,\replicaIdxAlt\rangle}{\{\dprfContribution{\replicaIdx}\}_{\replicaIdx\in\replicaIdxSubset}}$
  \label{line:vssRecoverNew:dprfEval}
  \State $\vssShareNew{\replicaIdxAlt} \gets \langle (\vssSecret(\replicaIdxAlt) - \maskingPolyY{\replicaIdxAlt}), \vssNoShare, \ldots, \vssNoShare\rangle$
  \label{line:vssRecoverNew:vssShare}
  \If{$\vssVerifyNew{\vssPubKeyNew{\replicaIdxAlt}}{\vssCommitmentNew}{\nonce}{\vssShareNew{\replicaIdxAlt}} = \textrm{false}$}
  \label{line:vssRecoverNew:vssVerifyNew}
  \State \Return \vssFailure
  \EndIf
  
  \State \Return \vssShareNew{\replicaIdxAlt}
  \EndProcedure
\end{algorithmic}
\end{minipage}
\caption{Pseudocode for our verifiable secret sharing scheme}
\label{fig:pcode}
\end{figure*}

\vssShareSecretNew is modified to take in all of the private keys
$\{\dprfPrivKey{\replicaIdx}\}_{\replicaIdx\in\nats{\numreplicas}}$
for the DPRF \dprfScheme, as well as the other arguments included in
its definition in \secref{sec:recovery:vss}.  (For this reason,
our construction requires each dealer to have a distinct set of
parameters for its sharings, i.e., produced by its own call to
\vssInitNew.)  
This enables the dealer to evaluate \dprfScheme itself,
which it does on $\langle \nonce, \replicaIdx \rangle$ for each
$\replicaIdx \in \nats{\numreplicas}$
(lines~\ref{line:vssShareSecretNew:loop-dprfEval}--\ref{line:vssShareSecretNew:dprfEval}),
where \nonce is a new, random \vssSecParam-bit nonce
(line~\ref{line:vssShareSecretNew:nonce}).  The resulting values
$\{\maskingPolyY{\replicaIdx}\}_{\replicaIdx \in \nats{\numreplicas}}$
are divided into \badfraction groups of size $\thresh-1$, each group
being used to construct a set of $\thresh-1$ points
$\maskingPolyPoints{\maskingPolyIdx} \gets \{\langle\replicaIdx,
\maskingPolyY{\replicaIdx}\rangle \mid (\maskingPolyIdx-1)(\thresh-1)
< \replicaIdx \le \maskingPolyIdx(\thresh-1)\}$
(line~\ref{line:vssShareSecretNew:maskingPolyPoints}) on which
\vssMakePoly is invoked
(line~\ref{line:vssShareSecretNew:vssMakePoly}).  The resulting
$\maskingPoly{\maskingPolyIdx} \in \polynomials{\vssModulus}$ is then
shared using \vssScheme
(line~\ref{line:vssShareSecretNew:vssShareDPRFVal}).  Recall that by
the definition of \vssMakePoly, each
\vssShareNew{\replicaIdx}{\maskingPolyIdx} thus produced satisfies
$\vssShareNew{\replicaIdx}{\maskingPolyIdx} =
\maskingPolyY{\replicaIdx}$.  Of course, the input secret \vssSecret
is also shared (line~\ref{line:vssShareSecretNew:vssShareVSSVal}).
The results of these sharings are grouped according to replica index
\replicaIdx and returned as \vssShareNew{\replicaIdx} for each
$\replicaIdx \in \nats{\numreplicas}$, along with all of the sharing
commitments \vssCommitmentNew and the nonce \nonce
(line~\ref{line:vssShareSecretNew:return}).

\vssVerifyNew and \vssReconstructNew operate in the natural way.
\vssVerifyNew verifies the commitment \vssCommitmentNew{0} and share
\vssShareNew{\replicaIdx}{0}
(line~\ref{line:vssVerifyNew:vssVerifyVSSShare}) produced in the
sharing of \vssSecret, as well as verifying the commitment
\vssCommitmentNew{\maskingPolyIdx} and share
\vssShareNew{\replicaIdx}{\maskingPolyIdx}
(line~\ref{line:vssVerifyNew:vssVerifyMaskingPolyShare}) produced in
the sharing of \maskingPoly{\maskingPolyIdx}.  In addition, it
verifies (intuitively) that
$\vssShareNew{\replicaIdx}{\maskingPolyIdx} =
\maskingPolyY{\replicaIdx}$ (line~\ref{line:vssVerifyNew:dprfVerify}).
The latter two verifications are skipped if
$\vssShareNew{\replicaIdx}{1} = \vssNoShare$
(line~\ref{line:vssVerifyNew:testRecovered}), which occurs if the
share \vssShareNew{\replicaIdx} was recovered (see below).  In this
case, $\vssShareNew{\replicaIdx}{\maskingPolyIdx} = \vssNoShare$ for
all $\maskingPolyIdx \in \nats{\badfraction}$ (or should be, and so
any $\maskingPolyIdx \in \nats{\badfraction}$ for which
$\vssShareNew{\replicaIdx}{\maskingPolyIdx} \neq \vssNoShare$ is just
ignored).  \vssReconstructNew simply uses \vssVerifyNew to verify each
share \vssShareNew{\replicaIdx} provided as input
(line~\ref{line:vssReconstructNew:vssVerifyNew}) and then submits
\vssCommitmentNew{0} and the inputs $\{\langle
\vssPubKey{\replicaIdx}, \vssShareNew{\replicaIdx}{0}
\rangle\}_{\replicaIdx \in \replicaIdxSubset}$ to \vssReconstruct to
reconstruct \vssSecret
(line~\ref{line:vssReconstructNew:vssReconstruct}).

\vssRecoverContribNew{\vssCommitmentNew}{\nonce}{\vssPrivKeyNew{\replicaIdx}}{\vssShareNew{\replicaIdx}}{\replicaIdxAlt}
is invoked at replica \replicaIdx to construct its contribution to
enable replica \replicaIdxAlt to reconstruct its share
\vssShareNew{\replicaIdxAlt}.  \vssRecoverContribNew returns
\vssShareNew{\replicaIdx}{0} blinded by
\vssShareNew{\replicaIdx}{\maskingPolyIdx}
(line~\ref{line:vssRecoverContribNew:return}) where $\maskingPolyIdx
\gets \lceil \replicaIdxAlt/(\thresh-1)\rceil$.  Then, so that
replica \replicaIdxAlt can recover its share of the original secret,
replica \replicaIdx also returns its share of the DPRF scheme
\dprfScheme evaluated at $\langle \nonce, \replicaIdxAlt \rangle$
(line~\ref{line:vssRecoverContribNew:dprfContrib}).

\vssRecoverVerifyNew{\vssCommitmentNew}{\nonce}{\vssRecoveryShareNew{\replicaIdx}}{\vssPubKeyNew{\replicaIdx}}{\replicaIdxAlt}
is executed by replica \replicaIdxAlt to verify that replica
\replicaIdx performed \vssRecoverContribNew correctly.  The output of
\vssRecoverContribNew contributed by replica \replicaIdx is passed
into \vssRecoverVerifyNew as \vssRecoveryShareNew{\replicaIdx} and is
parsed into its constituent components in
line~\ref{line:vssRecoverVerifyNew:parseVssRecoveryShareNew}.  First,
the DPRF contribution \dprfContribution{\replicaIdx} is checked on
line~\ref{line:vssRecoverVerifyNew:dprfVerify} to ensure that it
corresponds to a correct evaluation of the DPRF scheme \dprfScheme at
the point $\langle \nonce, \replicaIdxAlt\rangle$.
\vssRecoverVerifyNew then combines the commitments
(line~\ref{line:vssRecoverVerifyNew:vssCombineCommitments}) and uses
\vssVerify (line~\ref{line:vssRecoverVerifyNew:vssVerify}) to check
that the blinded share \vssBlindedShare was created correctly.  If
both checks pass, then \vssRecoverVerifyNew returns true.

\vssRecoverNew{\vssCommitmentNew}{\nonce}{\{\langle
  \vssPubKeyNew{\replicaIdx},
  \vssRecoveryShareNew{\replicaIdx}\rangle\}_{\replicaIdx\in\replicaIdxSubset}}{\replicaIdxAlt}{\vssPubKeyNew{\replicaIdxAlt}}
is executed at replica \replicaIdxAlt to recover its share
\vssShareNew{\replicaIdxAlt}.  In particular,
\vssShareNew{\replicaIdxAlt}{0} will be a share of the original
polynomial for \replicaIdxAlt.  \vssRecoverNew first invokes
\vssRecoverVerifyNew to make sure that the share sent by each replica
$\replicaIdx \in \replicaIdxSubset$ is correct
(line~\ref{line:vssRecoverNew:vssRecoverVerifyNew}).  \vssRecoverNew
then leverages \vssReconstruct
(line~\ref{line:vssRecoverNew:vssReconstruct}) to reconstruct a
polynomial $\vssSecret \in \polynomials{\vssModulus}$ that is the sum
of the polynomial originally shared in \vssShareSecretNew that
resulted in commitment \vssCommitmentNew{0} and the \maskingPolyIdx-th
masking polynomial \maskingPoly{\maskingPolyIdx} that resulted in
commitment \vssCommitmentNew{\maskingPolyIdx}, where $\maskingPolyIdx
= \lceil \replicaIdxAlt/(\thresh-1)\rceil$.  \vssRecoverNew then
evaluates $\vssSecret(\replicaIdxAlt)$ and subtracts
$\maskingPoly{\maskingPolyIdx}(\replicaIdxAlt) =
\dprfEval{\langle\nonce,\replicaIdxAlt\rangle}{\{\dprfContribution{\replicaIdx}\}_{\replicaIdx\in\replicaIdxSubset}}$
(lines~\ref{line:vssRecoverNew:vssReconstruct}--\ref{line:vssRecoverNew:vssShare})
to obtain \vssShareNew{\replicaIdxAlt}{0}.

\subsection{Security}
\label{sec:recovery:security}

Below, we prove that our modified VSS scheme still satisfies the security properties guaranteed by the underlying VSS protocol.

\paragraph{Hiding}
Suppose that the underlying DPRF scheme \dprfScheme and VSS scheme
\vssScheme are secure, and let \vssHidingAdversary{\vssSchemeNew} be a
hiding adversary for \vssSchemeNew.  We claim that if
\vssHidingAdversary{\vssSchemeNew} is legitimate, then for any
commitment \vssCommitmentNew, the set of indices \replicaIdxSubset for
which \vssHidingAdversary{\vssSchemeNew} obtains the shares
$\{\vssShareNew{\replicaIdx}{0}\}_{\replicaIdx \in \replicaIdxSubset}$
produced in line~\ref{line:vssShareSecretNew:vssShareVSSVal} (i.e., in
its invocation of \vssHidingOracle{\vssSchemeNew}{\vssHidingBit} that
returned \vssCommitmentNew) satisfies
$\cardinality{\replicaIdxSubset} < \thresh$.  To see why, note that
\vssHidingAdversary{\vssSchemeNew} can obtain
\vssShareNew{\replicaIdx}{0} for any \replicaIdx in one of three ways:
(i) by invoking
\vssReplicaOracle{\vssSchemeNew}{\replicaIdx}{\vssCompromiseCmd}; (ii)
by invoking
\vssReplicaOracle{\vssSchemeNew}{\replicaIdx}{\vssContribCmd{\vssCommitmentNew}};
or (iii) by invoking
\vssReplicaOracle{\vssSchemeNew}{\replicaIdxAlt}{\vssRecoverCmd{\vssCommitmentNew}{\replicaIdx}}
at each $\replicaIdxAlt \in \replicaIdxSubsetAlt$ where
$\cardinality{\replicaIdxSubsetAlt} \ge \thresh$, in which case
\vssHidingAdversary{\vssSchemeNew} can recover
\vssShareNew{\replicaIdx}{0} using the \vssRecoverNew routine
(line~\ref{line:vssRecoverNew:vssShare}).  Critically, invoking
\vssReplicaOracle{\vssSchemeNew}{\replicaIdxAlt}{\vssRecoverCmd{\vssCommitmentNew}{\replicaIdx}}
at each $\replicaIdxAlt \in \replicaIdxSubsetAlt$ where
$\cardinality{\replicaIdxSubsetAlt} < \thresh$ yields no useful
information about \vssShareNew{\replicaIdx}{0}, since when
$\cardinality{\replicaIdxSubsetAlt} < \thresh$, the value
\maskingPolyY{\replicaIdx} (line~\ref{line:vssRecoverNew:dprfEval})
and so $\vssShareNew{\replicaIdx}{0} = \vssSecret(\replicaIdx) -
\maskingPolyY{\replicaIdx}$
(line~\ref{line:vssRecoverNew:vssShare}) cannot be predicted
nonnegligibly better than guessing it at random, due to the security
of the DPRF \dprfScheme (i.e., (\ref{eqn:dprf})).  Because
\vssHidingAdversary{\vssSchemeNew} is legitimate, it thus obtains
\vssShareNew{\replicaIdx}{0} for only fewer than \thresh values of
\replicaIdx, and so by the security of \vssScheme, its success (in the
sense of (\ref{eqn:hiding})) is negligible in \vssSecParam.

\paragraph{Binding}
A binding adversary \vssBindingAdversary{\vssSchemeNew} is provided
inputs $\langle \vssModulus, \{\langle\vssPubKeyNew{\replicaIdx},
\vssPrivKeyNew{\replicaIdx}\rangle\}_{\replicaIdx \in
  \nats{\numreplicas}}\rangle \gets
\vssInitNew{1^{\vssSecParam}}{\thresh}{\numreplicas}$, and
\textit{succeeds} if it outputs \vssCommitmentNew, $\{
\vssShareNew{\replicaIdx}\}_{\replicaIdx \in
  \replicaIdxSubset}$ and $\{\vssShareNewAlt{\replicaIdx}\}_{\replicaIdx \in
  \replicaIdxSubsetAlt}$ for which
\[
\begin{array}{ll}
    & \vssReconstructNew{\vssCommitmentNew}{\nonce}{\{\langle \vssPubKeyNew{\replicaIdx}, \vssShareNew{\replicaIdx}\rangle\}_{\replicaIdx\in\replicaIdxSubset}} = \vssSecret \\
    \wedge & \vssReconstructNew{\vssCommitmentNew}{\nonce}{\{\langle \vssPubKeyNew{\replicaIdx}, \vssShareNewAlt{\replicaIdx}\rangle\}_{\replicaIdx\in\replicaIdxSubsetAlt}} = \vssSecretAlt \\
    \wedge & \vssSecret \neq \vssFailure \wedge \vssSecretAlt \neq \vssFailure \wedge \vssSecret \neq \vssSecretAlt
    \end{array}
\]
Let \vssSecret and \vssSecretAlt be values satisfying this condition.
Then,
\begin{align*}
\vssSecret & = \vssReconstruct{\vssCommitmentNew{0}}{\{\langle
    \vssPubKey{\replicaIdx}, \vssShareNew{\replicaIdx}{0}\rangle\}_{\replicaIdx\in\replicaIdxSubset}} \\
\vssSecretAlt & = \vssReconstruct{\vssCommitmentNew{0}}{\{\langle
    \vssPubKey{\replicaIdx}, \vssShareNew{\replicaIdx}{0}\rangle\}_{\replicaIdx\in\replicaIdxSubsetAlt}}
\end{align*}
where $\langle \vssModulus, \{\langle\vssPubKey{\replicaIdx},
\vssPrivKey{\replicaIdx}\rangle\}_{\replicaIdx \in
  \nats{\numreplicas}}\rangle \gets
\vssInit{1^{\vssSecParam}}{\thresh}{\numreplicas}$ (see
lines~\ref{line:vssInitNew:vssInit}
and~\ref{line:vssReconstructNew:vssReconstruct}).  That is, breaking
binding for \vssSchemeNew implies breaking binding for \vssScheme, and
so if \vssScheme ensures the binding property, then so does
\vssSchemeNew.

\section{A Private, Byzantine Fault-Tolerant Key-Value Store}
\label{sec:pbft}

In this section we describe how to incorporate \ourname{} into
PBFT~\cite{castro2002practical}. PBFT works with arbitrary applications,
but for simplicity, we will use a simple application that captures
how secret values are handled in our PBFT extension: a private, replicated 
key/value service that
tolerates Byzantine faults of replicas. The service provides
two APIs, {\sc put}(k, v) and {\sc get}(k), by which a client can write a value to a key or read a value
previously written to a key. The correctness of the value read from a
key is ensured despite up to \numfaulty Byzantine-faulty replicas, and
values are also kept private from \numfaulty faulty replicas, using
our verifiable secret-sharing approach.  Similar to previous works
(e.g.,~\cite{cachin2005constantinople,miller2016honeybadger}), a
client shares a secret value directly among the replicas, and a
consensus protocol drives agreement on a verifiable digest of the
value.

Our design assumes a classical asynchronous Byzantine model.
Specifically, we have $\numreplicas{} = 3\numfaulty{} + 1$ replicas,
$\le\numfaulty{}$ of which are Byzantine.  Any entity that sends a
request into the system is a client and in particular, clients and
replicas may not be distinct.  Any client may be Byzantine, but our
guarantees are only defined with respect to honest clients.  The
network is assumed to be asynchronous, but will eventually go through
periods of synchrony in which messages are delivered within a known
time bound and correct replicas and clients make progress at a known
rate.  We assume that each message is signed by its sender so that
its origin is known, subject to standard cryptographic assumptions.

We represent the state of the replicated service as a key-value store.
Every client in the system is allowed to view all keys in the store.
However, the service maintains a (potentially dynamic) access control
policy that specifies for every client the values it is allowed to
open.  Under these assumptions, we provide the standard guarantees
provided by a Byzantine fault tolerant protocol:

\begin{itemize}
\item \textbf{Linearizability}~\cite{herlihy1990linearizability}. If a
  client sends a request to the replicated service, then the service's
  response is consistent with an execution where the client's request
  was executed instantaneously at some point between when the request
  was sent and the response was received.
\item \textbf{Liveness}. If the network is synchronous, then every
  client request will get a response.
\end{itemize}

In addition to these standard properties, our design offers the
following privacy property:

\begin{itemize}
\item \textbf{Privacy}.  A value written to a key by a correct client
  where the access-control policy prohibits access by any faulty
  client, remains hidden from \numfaulty Byzantine servers.
\end{itemize}

\subsection{Setup and Log}

In addition to setting up authenticated communication channels among all
parties, 
in a setup phase, \vssInitNew is called for every client in the system and is
  part of the public/private key infrastructure.
The client takes the role of the dealer in \vssInitNew while each replica takes
  the role of a participant.
In particular, each client knows the secret key for all replicas corresponding to its
  invocation of \vssInitNew.

Every replica stores a full copy of the K-V store. For each key there are two
value entries, a public value (keyed $K$-pub) and a private value (keyed
$K$-priv). A replica maintains a bounded log of pending commands, waiting to be
committed. The size of the log cannot grow beyond a certain system-wide
parameter $W$. Once an entry in the log becomes committed, it is applied to the
K-V store. 
No replica or leader starts handling slot $j+W$ in the sequence of commands before it learns
that $2f+1$ replicas have committed all commands up to $j$. When a replica
learns that all commands up to $j$ have been committed by $2f+1$ replicas, it
evicts them from the log.

\subsection{Views}

Our solution employs a classical framework~\cite{dwork1988consensus,castro2002practical} that revolves around an
explicit ranking among proposals via \emph{view} numbers.
Replicas all start with an initial view, and progress from one view to the next.
They accept requests and respond to messages only in their current view.

In each view there is a single designated \textit{leader}.  
In a view, zero or more decisions may be reached.
This strategy separates safety from liveness: It maintains safety even if the
system exhibits arbitrary communication delays and again up to $f$ Byzantine
failures; it provides progress during periods of synchrony.

If a sufficient number of replicas suspect that the leader is faulty, then a
view change occurs and a new leader is elected.
The mechanism to trigger moving to a higher view is of no significance for
safety, but it is crucial for liveness. 

\subsection{Common Mode Protocol}

A client \putkey is split into two parts, public and private. The public part
is concerned with setting sequence ordering of requests. 
The private part stores a private value. 

More specifically, in a \putkey{K}{V} request, the client privately shares $V$ by via
\vssShareSecretNew, and sends the corresponding shares to every replica in the system.
The public part of \putkey{K}{V} consists of 
a client sending a \putkey{K}{c_V} request to the current leader. $c_V$ is
a global commitment to the polynomial
\vssSecret that binds the shares of each replica as a verifiable valid
share of \vssSecret).

The leader waits until its local log has length $< W$. It then 
extends its local log with the \putkey request, and sends a
pre-prepare (ordering-request) containing its log tail. 
We discuss below the protocol for a leader to pick an
initial log when starting a new view. 

A replica \textit{accepts} a pre-prepare from the leader of the current
view if it has valid format, if it extends
any previous pre-prepare from this leader,
if its log has fewer than $W$ pending entries,
and if the replica received a valid share corresponding to $c$. 
If the leader pre-prepare message has a valid format, but the replica did
not receive the corresponding share for it, it starts a timer for share-recovery 
(see more below). 

Upon accepting a pre-prepare, a replica extends its local log to include the new
request and broadcasts a \emph{prepare} message to all replicas that includes
the new log tail.
Replicas wait to collect a \emph{commit-certificate}, a set of $2f+1$ prepare responses for the current log tail. 
Then the replica broadcasts a \emph{commit} message carrying the commit-certificate to
the other replicas. 
A decision is reached in a view on a new log tail 
when $2f+1$ distinct replica have sent a commit message for it. 

\begin{figure}
\begin{subfigure}[b]{1\linewidth}
    \includegraphics[trim={0 2in 0in 2in}, width=1\linewidth]{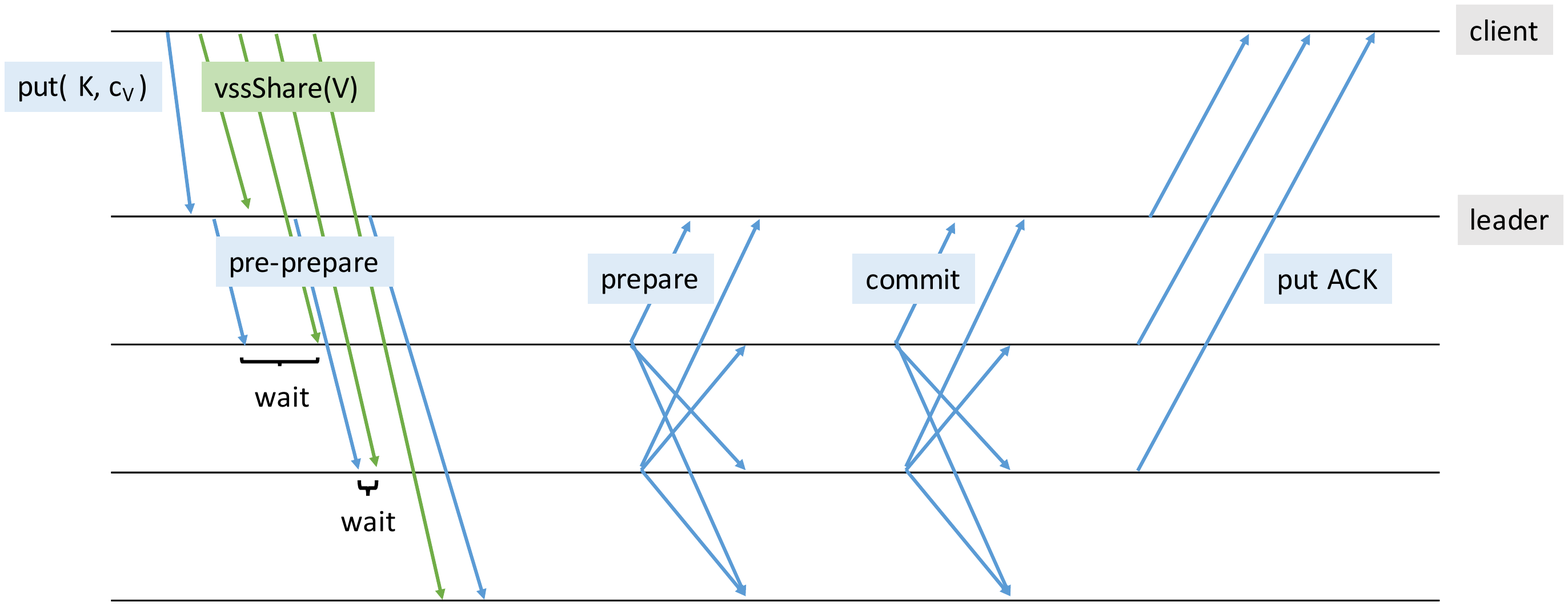}
    \caption{faultless IO path}
    \label{fig:pbft:sec}
\end{subfigure}
\begin{subfigure}[b]{1\linewidth}
    \includegraphics[trim={0 2in 0in 2in}, width=1\linewidth]{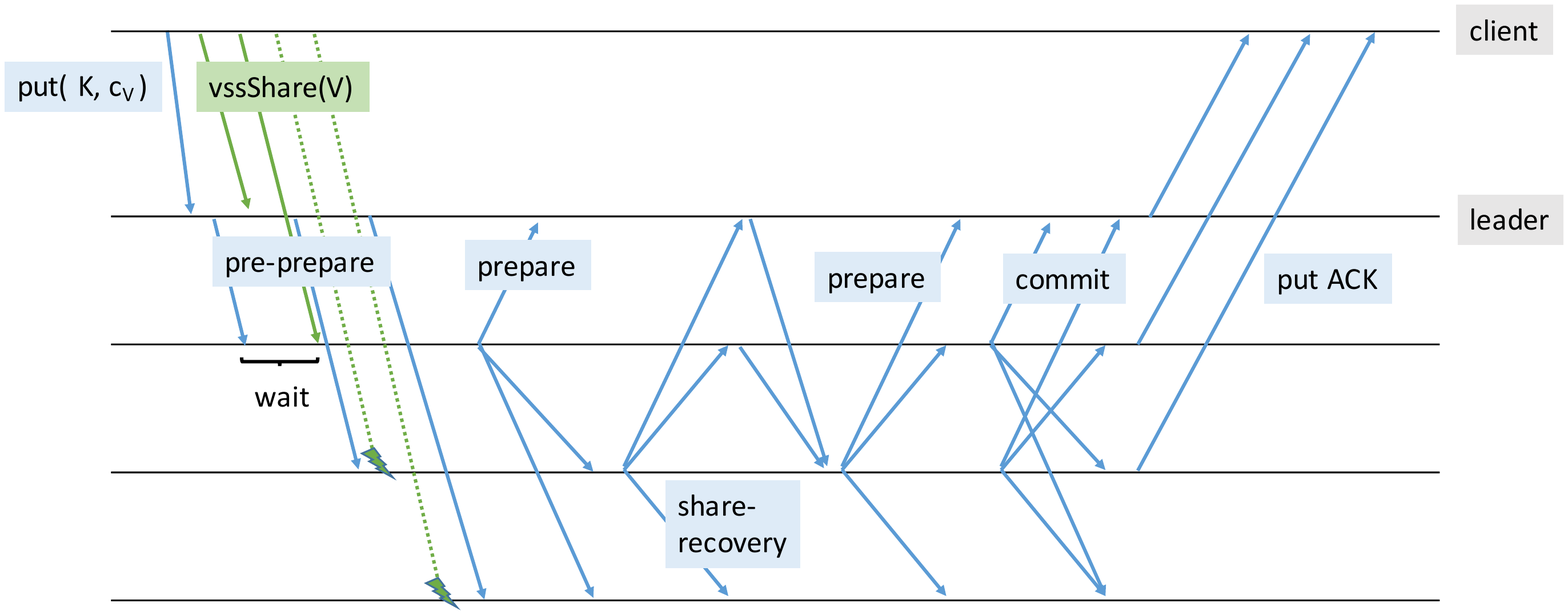}
    \caption{IO path with recovery}
    \label{fig:pbft:recovery}
\end{subfigure}
\caption{\putkey commmon mode}\label{fig:changeprefer}
\label{fig:pbft}
\end{figure}

When a replica learns that a \putkey{K}{V} request has been committed to the log,
it inserts to its local key-value store two entries, a global entry $(( K || public),
c_V)$ containing the global commitment to $V$, and a private entry $((K ||
private), share)$ containing the replica's private share. 
The replica then responds to the client with an \putkey acknowledgement message
containing $K$ and $c$. 
A client waits to receive $2f+1$ \putkey responses to complete the request.
\figref{fig:pbft:sec} depicts the \putkey io path, and 
\figref{fig:pbft:recovery} the \putkey io path when shares are missed.

The client \getkey{K} protocol consists of sending the \getkey request to
the current leader. The pre-prepare, prepare and commit phases of the ordering
protocol are carried as above, without the need to wait for shares. At the final
stage, when a replica executes the \getkey requests, it returns its share to the
client in a response. If the replica is missing its share, it initiates the share-recovery
protocol. The client waits to receive $f+1$ valid \getkey responses.
It uses \vssVerifyNew to verify each response, and \vssReconstructNew to
reconstruct the secret value from the responses. 

\subsection{Share-Recovery Protocol}

There are several circumstances in the protocol when 
a replica discovers it is missing its private share of a request and needs to
recover it (see above). 
To initiate share-recovery, a replica broadcasts a recovery request.
Other replicas respond to a share-recovery request with
the output of \vssRecoverContribNew.
After receiving a response, the original replica uses \vssRecoverVerifyNew to check the response.
If the response is valid, it is stored and if it is invalid, it is dropped.
When it receives $\numfaulty + 1$ valid responses, the replica uses \vssRecoverNew
  to recover its missing secret share.

\subsection{Common mode performance}

The common mode protocol incurs the following performance costs. 
The client interaction with the BFT replicated service is linear, since it needs
to populate all replicas with shares.
Additionally, the client collects $f+1$ responses from servers. 

The communication among the replicas to achieve an ordering decision is
quadratic. There are several practical variants of BFT replication that achieve linear
communication 
during periods of synchrony and when a leader is non-faulty
(e.g.,~\cite{martin2006fab,kotla2009zyzzyva,golan-gueta2018sbft})
These improvements are left outside the scope of this paper, however, our linear
AVSS protocol is designed so it can be incorporated within them 
without increasing the asymptotic complexity of the common mode.

In terms of latency, the sharing protocol is non-interactive and single-round,
hence it can be performed concurrently with the leader broadcast.
Recovery incurs extra latency since each replica must ask at least $\numfaulty + 1$ 
  correct replicas for their contribution.
In the original BFT protocol, recovering a missing request only requires asking
  $1$ correct replica for the request data.
In both cases, the recovery protocol is interactive and single-round, so there are
  no asymptotic increases in latency.
However, in practice, there will be a difference in latency between the two scenarios.

\medskip

\emph{Remaining Items.} For completeness, we describe the view change and state transfer
protocols in \appref{sec:app:protocoldescription}.
A proof of linearizability, liveness and privacy appears in \appref{sec:app:pbftproof}.


\section{Instantiation and Implementation}
\label{sec:impl}

We first see how to instantiate our distributed pseudorandom function
using the techniques presented in Naor, et
al.~\cite{naor1999distributed}.  We then cover two ways to instantiate
our verifiable scheme, using Pedersen secret
sharing~\cite{pedersen1991non-interactive} and one from Kate,
et al.~\cite{kate2010constant-size}.  Finally, we go over
implementation details and the programming API that we expose for
application developers.

\subsection{Distributed Pseudorandom Function Instantiation}
Our distributed pseudorandom function \dprfScheme consists of four
algorithms: \dprfInit, \dprfContrib, \dprfVerify, and \dprfEval.  Our
implementation defines them as follows:

\begin{compactitem}
\item 
$\dprfInit{1^{\dprfSecParam}}{\thresh}{\numreplicas}{\dprfDomain}{\residues{\vssModulus}}$,
  first chooses a generator \groupElem{} of \group{} of order $\vssModulus$.
  A \thresh out of \numreplicas secret sharing of a private value $\dprfExp \in 
  \residues{\vssModulus}$
  is produced using Shamir secret sharing~\cite{shamir1979how}, of
  which the shares are $\{\dprfExpShare{\replicaIdx}\}_{\replicaIdx
    \in \nats{\numreplicas}}$.  \dprfPubKey{\replicaIdxAlt} is set to
  $\langle \groupElem{}, \groupElem{}^{\dprfExp},
  \{\groupElem{}^{\dprfExpShare{\replicaIdx}}\}_{\replicaIdx \in
    \nats{\numreplicas}} \rangle$ for all $\replicaIdxAlt =
  1..\numreplicas$.  \dprfPrivKey{\replicaIdxAlt} is set to
  \dprfExpShare{\replicaIdxAlt} for all $\replicaIdxAlt =
  1..\numreplicas$.  \dprfInit outputs
  $\{\langle\dprfPubKey{\replicaIdx},
  \dprfPrivKey{\replicaIdx}\rangle\}_{\replicaIdx \in
    \nats{\numreplicas}}$.

\item
$\dprfContrib{\dprfPrivKey{\replicaIdx}}{\dprfInput}$ first computes
  $\dprfContribIdx{\replicaIdx} = \dprfHash(\dprfInput)^{\dprfExpShare{\replicaIdx}}$ where
  $\dprfHash: \{0,1\}^{\ast} \rightarrow \group{}$ is a hash function that is
  modeled as a random oracle.  Here, \dprfExpShare{\replicaIdx} is
  obtained from the \dprfPrivKey{\replicaIdx}.  Let \dprfRandom be a
  randomly generated element of \residues{\vssModulusQRs}.  Then, we
  let $\dprfChallenge{\replicaIdx} \gets
  \dprfZKPHash(\dprfHash(\dprfInput), \groupElem{},
  \dprfContribIdx{\replicaIdx},
  \groupElem{}^{\dprfExpShare{\replicaIdx{}}},
  \dprfHash(\dprfInput)^{\dprfRandom}, \groupElem{}^{\dprfRandom})$,
  where $\dprfZKPHash: \{0,1\}^{\ast} \rightarrow
  \residues{\vssModulus}$ is a hash function modeled as a random
  oracle.  We set $\dprfProof{\replicaIdx} \gets
  \dprfExpShare{\replicaIdx}\dprfChallenge{\replicaIdx} + \dprfRandom
  \bmod \vssModulus$.  \dprfContrib then outputs $\langle
  \dprfContribIdx{\replicaIdx}, \dprfProof{\replicaIdx},
  \dprfChallenge{\replicaIdx} \rangle$.

\item 
$\dprfVerify{\dprfPubKey{\replicaIdx}}{\dprfInput}{\dprfContribution}$
  first extracts \dprfContribIdx{\replicaIdx}, \dprfChallenge{\replicaIdx}, and \dprfProof{\replicaIdx} from 
  \dprfContribution.
Then, \groupElem{} and $\groupElem{}^{\dprfExpShare{\replicaIdx{}}}$ are extracted from
  \dprfPubKey{\replicaIdx}.
Finally, \dprfVerify returns true if $\dprfChallenge{\replicaIdx} = \dprfZKPHash(\dprfHash(\dprfInput), \groupElem{}, 
  \dprfContribIdx{i}, \groupElem{}^{\dprfExpShare{\replicaIdx{}}}, 
  H(\dprfInput)^{\dprfProof{\replicaIdx}}\dprfContribIdx{i}^{-\dprfChallenge{\replicaIdx}},
  \groupElem{}^{\dprfProof{\replicaIdx}}(\groupElem{}^{\dprfExpShare{\replicaIdx{}}})^{-\dprfChallenge{\replicaIdx}})$.

\item
$\dprfEval{\dprfInput}{\{\dprfContribution{\replicaIdx}\}_{\replicaIdx\in\replicaIdxSubset}}$
  first verifies each $\dprfContribution{\replicaIdx}$ using
  \dprfVerify.  If any of the verifications returns false, then
  \dprfEval returns \dprfFailure.  Otherwise, we extract
  $\dprfContribIdx{\replicaIdx}$ values from each
  \dprfContribution{\replicaIdx}.  Since the exponents of
  \dprfContribIdx{\replicaIdx} were shared in the exponent using
  Shamir secret sharing, \dprfEval uses Lagrange interpolation in the
  exponent to get the value of \dprfScheme at \dprfInput, hashes it into an
  element of \residues{\vssModulus} and ouputs that value.

\end{compactitem}

\subsection{Verifiable Secret Sharing Instantiations}
\label{sec:impl:vss}

For a verifiable secret sharing scheme, we require the functions
\vssInit, \vssShareSecret, \vssVerify, and \vssReconstruct.  To be
used in our construction, we require a few additional functions,
namely: \vssMakePoly and \vssCombineCommitments.  To support addition,
we simply require replicas to add the respective shares, by definition
of \vssCombineCommitments.  We now define all of these functions for
two secret sharing schemes.

\subsubsection{Pedersen Secret Sharing}
We describe how to fit the secret sharing scheme from
Pedersen~\cite{pedersen1991non-interactive} into our framework.

\begin{compactitem}
\item
$\vssInit{1^{\vssSecParam}}{\thresh}{\numreplicas}$ first chooses a
  safe prime $\vssModulus = 2\vssModulusQRs + 1$ at least \vssSecParam
  bits in length, for \vssModulusQRs a prime.  Also, we let \vssGen
  and \vssGenTwo be two distinct generators of the quadratic residues
  \qr{\residuesNonzero{\vssModulus}} of \residuesNonzero{\vssModulus}
  such that $\log_{\vssGen}(\vssGenTwo)$ is unknown.  Then,
  \vssPubKey{\replicaIdx} is set to $\langle \vssGen, \vssGenTwo
  \rangle$ for all \replicaIdx.  \vssPrivKey{\replicaIdx} is set to
  $\bot$ for all \replicaIdx.  The return value of \vssInit is
  $\langle \vssModulus, \{ \langle \vssPubKey{\replicaIdx},
  \vssPrivKey{\replicaIdx} \rangle \}_{\replicaIdx \in
    \nats{\numreplicas}} \rangle$.

\item
\vssShareSecret{\vssSecret}{\vssModulus}{\{\vssPubKey{\replicaIdx}\}_{\replicaIdx\in\nats{\numreplicas}}}
  first extracts the public key and gets \vssGen and \vssGenTwo as defined in \vssInit.
Set \vssSecret{\coeffIdx} be the coefficient of the $x^\coeffIdx$ term in \vssSecret and 
  $\vssSecret(\replicaIdx)$ be the evaluation of \vssSecret at point \replicaIdx.
Pick $\pedVssMasking \in \polynomials{\vssModulus}$ to be a random polynomial of degree 
  $\thresh - 1$.
Similarly, we let \pedVssMasking{\coeffIdx} be the coefficient of $x^\coeffIdx$ term in 
  \pedVssMasking and $\pedVssMasking(\replicaIdx)$ be the evaluation of \pedVssMasking at 
  point \replicaIdx.
Now, we set \vssShare{\replicaIdx} to be 
  $\langle \vssSecret(\replicaIdx), \pedVssMasking(\replicaIdx) \rangle$.
Here, all linear operations on $\vssShare{\replicaIdx}$ values are just done element-wise.
Set \vssCommitment to be $\{\vssGen^{\vssSecret{\coeffIdx}}\vssGenTwo^{\pedVssMasking{\coeffIdx}}
  \}_{\coeffIdx \in \nats{\thresh}}$.
Then, \vssShareSecret returns $\langle \vssCommitment, \{ \vssShare{\replicaIdx} \}_{\replicaIdx\in\nats{\numreplicas}} \rangle$.

\item 
\vssVerify{\vssPubKey{\replicaIdx}}{\vssCommitment}{\vssShare{\replicaIdx}}
  first extracts $\{\vssGen^{\vssSecret{\coeffIdx}}\vssGenTwo^{\pedVssMasking{\coeffIdx}}\}
    _{\coeffIdx \in \nats{\thresh}}$ from \vssCommitment.
Then, $\vssSecret(\replicaIdx), \pedVssMasking(\replicaIdx)$ is extracted from 
  \vssShare{\replicaIdx}.
We return true if $\vssGen^{\vssSecret(\replicaIdx)}\vssGenTwo^{\pedVssMasking(\replicaIdx)} = \prod_{\coeffIdx = 0}^{\thresh-1}
  (\vssGen^{\vssSecret{\coeffIdx}}\vssGenTwo^{\pedVssMasking{\coeffIdx}})^{\replicaIdx^\coeffIdx}$
  and false otherwise.

\item
\vssReconstruct{\vssCommitment}{\{\langle \vssPubKey{\replicaIdx}, \vssShare{\replicaIdx}\rangle\}_{\replicaIdx\in\replicaIdxSubset}} first calls
  $\vssVerify(\vssPubKey{\replicaIdx}, \vssCommitment, \vssShare{\replicaIdx})$ for all $\replicaIdx \in \replicaIdxSubset$.
If all of \vssVerify calls return true, then we continue.
Otherwise, \vssReconstruct returns $\bot$.
Then, we extract $\vssSecret(\replicaIdx), \pedVssMasking(\replicaIdx)$ from each 
  \vssShare{\replicaIdx}.
Finally, we simply do a Lagrange interpolation in order to identify the unique degree 
  $\thresh - 1$ polynomial in \polynomials{\vssModulus} that goes through the points $(\replicaIdx, \vssSecret(\replicaIdx))$ 
  for all $\replicaIdx \in \replicaIdxSubset$ and return that value.

\item
$\vssMakePoly{\vssModulus}{\{\langle \maskingPolyX{\replicaIdx}, 
  \maskingPolyY{\replicaIdx} \rangle\}_{\replicaIdx \in \replicaIdxSubset}}$ does a Lagrange
  interpolation in order to identify the unique degree $\thresh - 1$ polynomial in \polynomials{\vssModulus} that goes through
  $(\maskingPolyX{\replicaIdx}, \maskingPolyY{\replicaIdx})$ and returns that as $s$.

\item 
$\vssCombineCommitments{\vssCommitment}{\vssCommitmentAlt}$ extracts 
  $\{\vssGen^{\vssSecret{\coeffIdx}}\vssGenTwo^{\pedVssMasking{\coeffIdx}} 
  \}_{\coeffIdx \in \nats{\thresh}}$ from \vssCommitment and
  $\{\vssGen^{\vssSecretAlt{\coeffIdx}}\vssGenTwo^{\pedVssMaskingAlt{\coeffIdx}} 
  \}_{\coeffIdx \in \nats{\thresh}}$ from \vssCommitmentAlt.
Then, it returns $\vssCommitmentAltAlt = 
  \{(\vssGen^{\vssSecretAlt{\coeffIdx}}\vssGenTwo^{\pedVssMaskingAlt{\coeffIdx}})
  (\vssGen^{\vssSecret{\coeffIdx}}\vssGenTwo^{\pedVssMasking{\coeffIdx}})
  \}_{\coeffIdx \in \nats{\thresh}}$.

\end{compactitem}

\subsubsection{Kate et al.\ Secret Sharing}
\label{sec:impl:vss:kate}
We describe how to fit the secret sharing scheme from Kate et al.~\cite{kate2010constant-size} into our
  framework.
Note that this secret sharing scheme also has a \emph{witness}, which proves that a particular
  share is consistent with the polynomial commitment.
Witnesses are additively homomorphic as well and can be manipulated the same way as the shares
  can.
In particular, we can perform polynomial interpolation in order to take a set of $\numfaulty$ witnesses
  and obtain the witness for any other share.
Additionally, we only need to send a witness when we transmit the corresponding share.
Thus, witnesses only increase the communication overhead by a constant factor.
In the description below, we assume that we have the witness corresponding to each share.

\begin{compactitem}
\item
$\vssInit{1^{\vssSecParam}}{\thresh}{\numreplicas}$ first chooses a safe prime \vssModulus at least
  \vssSecParam bits in length.
Then, we initialize two groups of order \vssModulus: $\group{}$ and $\group{t}$ such that there
  exists a bilinear map $\blmap \colon{} \group{} \times \group{} \rightarrow \group{t}$.
We then generate a $\bltrap \in \residues{\vssModulus}$ and pick a generator $\vssGen \in \group{}$.
Set \vssPubKey{\replicaIdx} to be $\langle \group{}, \group{t}, \blmap, \vssGen, 
  \{\vssGen^{\bltrap^{\coeffIdx}}\}_{\coeffIdx \in \nats{\numfaulty}} \rangle$
  and \vssPrivKey{\replicaIdx} to be $\bot$ for all \replicaIdx.
Then, we delete $\bltrap$.
Finally, \vssInit returns $\langle \vssModulus, \{ \langle \vssPubKey{\replicaIdx}, 
  \vssPrivKey{\replicaIdx} \rangle \}_{\replicaIdx \in \nats{\numreplicas}} \rangle$.

\item
\vssShareSecret{\vssSecret}{\vssModulus}{\{\vssPubKey{\replicaIdx}\}_{\replicaIdx\in\nats{\numreplicas}}}
  first extracts the public key and gets \vssGen and 
  $\{\vssGen^{\bltrap^{\coeffIdx}}\}_{\coeffIdx \in \nats{\numfaulty}}$.
Let \vssSecret{\coeffIdx} be the coefficient of the $x^\coeffIdx$ term in \vssSecret and 
  $\vssSecret(\replicaIdx)$ be the evaluation of \vssSecret at point \replicaIdx.
We now compute $\vssGen^{\vssSecret(\bltrap)}$ by computing 
  $\prod_{\coeffIdx=0}^{\numfaulty} (\vssGen^{\bltrap^{\coeffIdx}})^{\vssSecret{\coeffIdx}}$
  and assign it to \vssCommitment.
Now, using polynomial division, we can compute the coefficients of 
  $\frac{\vssSecret(x) - \vssSecret(\replicaIdx)}{x-\replicaIdx}$, which will allow us to compute 
  $\vssGen^{\frac{\vssSecret(\bltrap) - \vssSecret(\replicaIdx)}{\bltrap-\replicaIdx}}$ which is the
  witness for \vssShare{\replicaIdx}.
We also set \vssShare{\replicaIdx} to be $\vssSecret(\replicaIdx)$.
Finally, \vssShareSecret returns $\langle \vssCommitment, \{\vssShare{\replicaIdx}\}_{\replicaIdx\in\nats{\numreplicas}} \rangle$.

\item 
\vssVerify{\vssPubKey{\replicaIdx}}{\vssCommitment}{\vssShare{\replicaIdx}}
  first extracts $\vssGen^{\vssSecret(\bltrap)}$ from \vssCommitment, 
  $\vssSecret(\replicaIdx)$ from \vssShare{\replicaIdx}, and $\blmap$,
  $\vssGen$, $\vssGen^{\bltrap}$ from \vssPubKey{\replicaIdx}.
We also have access to the value $\vssGen^{\frac{\vssSecret(\bltrap) - \vssSecret(\replicaIdx)}{\bltrap-\replicaIdx}}$
  since the witness for the share is transmitted along with the share.
Then, \vssVerify returns true if $\blmap{\vssGen^{\vssSecret(\bltrap)}}{\vssGen}$ equals
  $\blmap{\vssGen^{\frac{\vssSecret(\bltrap) - \vssSecret(\replicaIdx)}{\bltrap-\replicaIdx}}}{\frac{\vssGen^{\bltrap}}{\vssGen^{\replicaIdx}}}
  \blmap{\vssGen}{\vssGen}^{\vssSecret(\replicaIdx)}$
  and false otherwise.

\item
\vssReconstruct{\vssCommitment}{\{\langle \vssPubKey{\replicaIdx}, \vssShare{\replicaIdx}\rangle\}_{\replicaIdx\in\replicaIdxSubset}} first calls
  \vssVerify{\vssPubKey{\replicaIdx}}{\vssCommitment}{\vssShare{\replicaIdx}} 
  for all $\replicaIdx \in \replicaIdxSubset$.
If all of \vssVerify calls return true, then we continue.
Otherwise, \vssReconstruct returns $\bot$.
Then, similarly to the Pedersen scheme, we extract $\vssSecret(\replicaIdx)$ from each 
  \vssShare{\replicaIdx} and do Lagrange interpolation to identify the original polynomial
  and return that value.

\item
$\vssMakePoly{\vssModulus}{\{\langle \maskingPolyX{\replicaIdx}, 
  \maskingPolyY{\replicaIdx} \rangle\}_{\replicaIdx \in \replicaIdxSubset}}$ works the same
  way as it does in the Pedersen scheme above.

\item 
$\vssCombineCommitments{\vssCommitment}{\vssCommitmentAlt}$ first extracts
  $\vssGen^{\vssSecret(\bltrap)}$ from \vssCommitment and $\vssGen^{\vssSecretAlt(\bltrap)}$ from 
  \vssCommitmentAlt.
We then set \vssCommitmentAltAlt to $(\vssGen^{\vssSecret{\bltrap}})(\vssGen^{\vssSecretAlt(\bltrap)})$
  and return that value.

\end{compactitem}

\subsection{Implementation}

We implement a secret shared BFT engine by layering PBFT~\cite{castro2002practical} with our secret sharing scheme.
Our implementation consists of 4700 lines of Python and 4800 lines of C.
We optimize our design for multicore environments, with one network thread running on a core which
  never blocks.
Additionally, we use one thread for every other core in order to do all cryptographic operations
  that are required by PBFT and our secret sharing scheme.
We use elliptic curve signatures with the secp256k1 library for all
  signature checking operations and the Relic library~\cite{relic} for all other cryptographic
  operations related to our scheme.
We also make a few optimizations for the Kate and Pedersen secret sharing schemes in order to make
  them faster.

\paragraph{Kate et al.} Kate et al.'s secret sharing scheme lends itself for extensive caching during setup time.
Once the powers $\vssGen^{\bltrap^{\coeffIdx}}$ are known for all $\coeffIdx$, we construct
  precomputation tables for each coefficient so that all exponentiations during runtime leverage
  these tables for efficiency.
In the sharing step, we first use the well known Horner's method to optimize the share evaluation.
However, we also note that each intermediate value obtained in Horner's method when evaluating
  $\vssSecret(\replicaIdx)$ is also the coefficient of the quotient polynomial 
  $\frac{\vssSecret(x) - \vssSecret(\replicaIdx)}{x-\replicaIdx}$ which means that we can do the
  necessary division required for free before using our precomputation tables to evaluate the
  quotient at $\bltrap$.
In the share verification step, we note that every verification requires the value of
  $\blmap{\vssGen}{\vssGen}$ so we can precompute that as well to save a bilinear map operation.
Also in the share verification phase, we note that the division of 
  $\frac{\vssGen^{\bltrap}}{\vssGen^{\replicaIdx}}$ only has $\numreplicas$ possible values which
  means that we can precompute all of these values as well.
Finally, when doing Lagrange interpolation, we know that the indices range from $0$ to 
  $\numreplicas-1$ and in the denominator, we need to compute the produce of differences of these
  indices.
Thus, to avoid taking inverses, we simply take inverses of all $\numreplicas$ values of the 
  differences which means that during runtime, we only have to do multiplications.

\paragraph{Pedersen} Pedersen's secret sharing scheme does not lend itself to as much caching since
  most of the values are unknown beforehand.
However, we do generate precomputation tables for both $\vssGen$ and $\vssGenTwo$ during setup and
  compute the inverses to make Lagrange interpolation easier.

\section{Evaluation}

Our evaluation seeks to answer two basic questions. First, we
investigate the costs of each API call in our secret sharing scheme.
Then, we look at how expensive it is to incorporate our secret sharing
scheme into a Byzantine Fault Tolerant key value store.  We
instantiate \ourname{} using the DPRF in Naor et al.~\cite{naor1999distributed} and
two different VSS schemes: Pedersen's VSS scheme~\cite{pedersen1991non-interactive} 
and Kate et al.'s VSS scheme~\cite{kate2010constant-size}.
We call the Pedersen instantiation Ped-\ourname{}
and the Kate et al.\ instantiation KZG-\ourname{}. Note that the
latter scheme has constant overhead on the replicas per sharing, while
Ped~\ourname{} has linear overhead on the replicas but with cheaper cryptographic
operations. We instantiate the BFT algorithm using PBFT~\cite{castro2002practical} 
and build a Byzantine Fault Tolerant key value store with secret shared state.

Our implementation uses the Relic~\cite{relic} cryptographic library
and, for our elliptic-curve algorithms, the \texttt{BN\_P254} curve.

\subsection{Microbenchmarks}

For our microbenchmarks, we evaluate each function in our full
asynchronous verifiable secret sharing scheme. We vary the number
of replicas from $4$ to $211$ and measure the latency and throughput
of each operation. We use EC2 \texttt{c5.xlarge} instances in order
to run our microbenchmarks, which have $4$ virtual CPUs per instance.

The module that implements our secret sharing scheme optimizes for
throughput, while compromising slightly on latency.  Each API call
runs on a single core; the task is run to completion and the result is
returned in the order that the tasks were enqueud. This maximizes
for throughput due to the lack of cross core communication, but at the
expense of request latency as many of the underlying cryptographic
operations can leverage multicore environments to execute faster.

Each microbenchmark ran for at least $60$ seconds and collected at least $30$
samples. Before computing the final statistic, we ignored any requests that
were completed in the first $10$ seconds and the last $10$ seconds of the run.
We report the aggregate throughput during the run and the mean and standard
deviation of the latency of each request completed in our run.

\subsubsection{\vssShareSecretNew Microbenchmark}

\begin{figure}
\begin{subfigure}[t]{0.475\columnwidth}
  \includegraphics[width=1\linewidth]{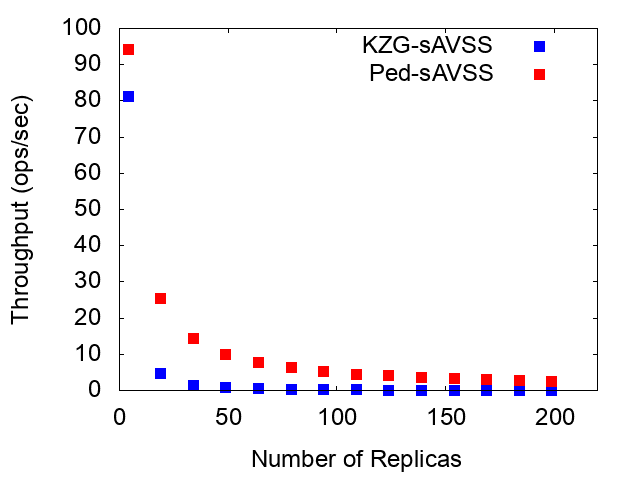}
  \caption{\vssShareSecretNew throughput}
  \label{fig:sharing:xput}
\end{subfigure}
\hfill
\begin{subfigure}[t]{0.475\columnwidth}
  \includegraphics[width=1\linewidth]{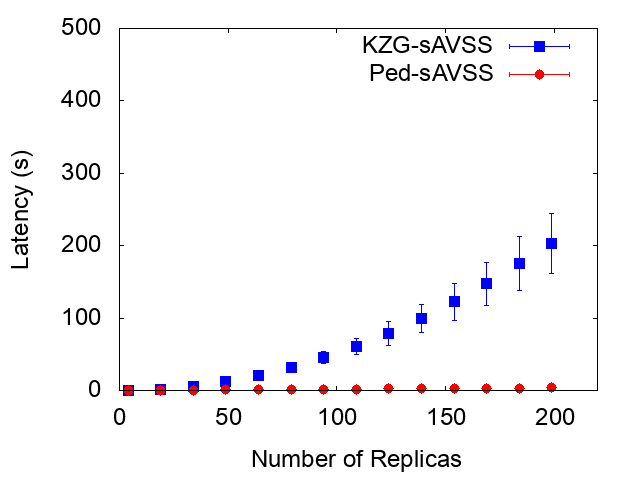}
  \caption{\vssShareSecretNew latency}
  \label{fig:sharing:latency}
\end{subfigure}
\caption{\vssShareSecretNew performance with varying \numreplicas}
\label{fig:sharing}
\end{figure}

\figref{fig:sharing:xput} shows that Ped-\ourname{} can sustain more
sharings per second than KZG-\ourname{} for all cluster sizes.
At $\numreplicas=4$, Ped-\ourname{} does $1.2$ times more sharings per second
than KZG-\ourname{}. However, this difference increases as the cluster size
increases, with Ped-\ourname{} sustaining $58$ times more sharings per second
than KZG-\ourname{} at $\numreplicas = 199$.

\begin{wrapfigure}{r}{0.485\columnwidth}
  \includegraphics[width=1\linewidth]{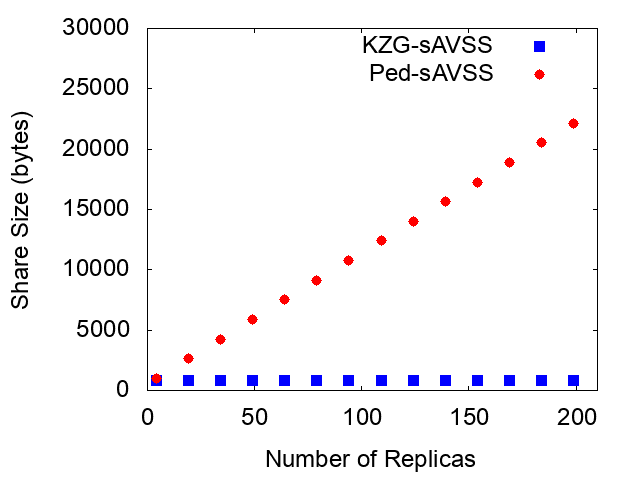}
  \caption{Volume the dealer transmits per replica to share a secret among
    \numreplicas replicas}
  \label{fig:sharing:bw}
\end{wrapfigure}

This performance difference between Ped-\ourname{} and KZG-\ourname{} is due to the 
underlying VSS scheme. KZG-\ourname{} computes witnesses for
each share, which involves evaluating a polynomial in the elliptic curve group.
Additionally, we see that the throughput decrease is quadratic
since evaluating each share (or witness) takes $O(\numreplicas)$ CPU time and there are
$\numreplicas$ shares so \vssShareSecretNew takes $O(\numreplicas^2)$ time for both
KZG-\ourname{} and Ped-\ourname{}. For the above reasons, we also see that 
KZG-\ourname{} has a higher latency than Ped-\ourname{} as we see in 
\figref{fig:sharing:latency}. This discrepancy also increases as the size of the cluster
increases.

\figref{fig:sharing:bw} shows the size of the share and associated metadata
that is sent to each replica after the client computes a share, which is equal
to the disk space that the replica needs to store a secret shared value. Since Ped-\ourname{}
has a linear overhead per sharing to each replica, we see the bandwidth and storage footprint increasing
linearly with the cluster size. Ped-\ourname{} requires about $1 KB$ of storage at $\numreplicas=4$
and increases to $23 KB$ at $\numreplicas = 211$.
Meanwhile, KZG-\ourname{} only requires each replica to store $860$ bytes of information
irrespective of the cluster size.
Note that, in both instances, we are storing a single $254$ bit integer.

\subsubsection{\vssVerifyNew Microbenchmark}

\begin{figure}
\begin{subfigure}[t]{0.475\columnwidth}
  \includegraphics[width=1\linewidth]{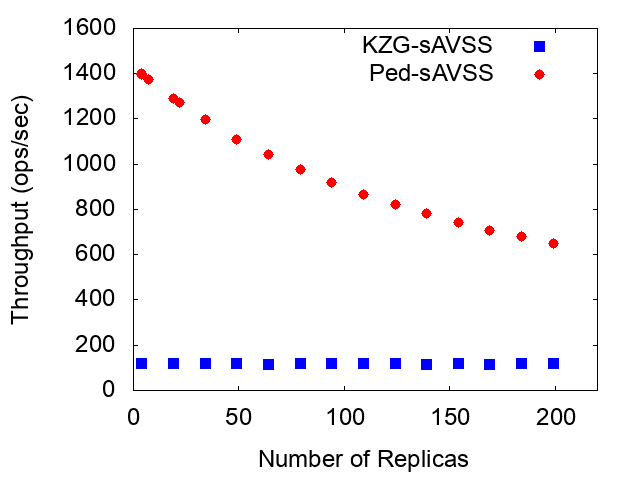}
  \caption{\vssVerifyNew throughput}
  \label{fig:shareverify:xput}
\end{subfigure}
\hfill
\begin{subfigure}[t]{0.475\columnwidth}
  \includegraphics[width=1\linewidth]{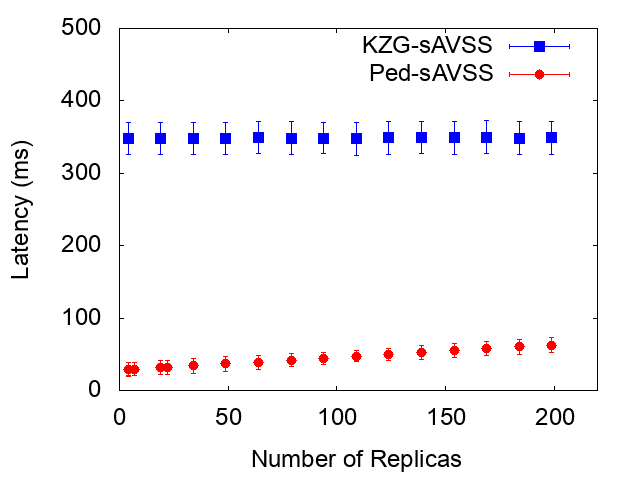}
  \caption{\vssVerifyNew latency}
  \label{fig:shareverify:latency}
\end{subfigure}
\caption{\vssVerifyNew performance with varying \numreplicas}
\label{fig:shareverify}
\end{figure}

\figref{fig:shareverify} shows that the throughput and latency of verifying a share,
which is done by the replicas upon receiving a share. We see that at $\numreplicas = 4$, 
Ped-\ourname{} has $12$ times higher throughput and $38$ times lower latency. 
At $\numreplicas = 211$, Ped-\ourname{} only outperforms KZG-\ourname{} by a factor of $5.4$ on
throughput and is only $5.4$ times faster. We also see another trend: KZG-\ourname{}'s 
latency and throughput stays constant at $117$ operations per second with a $350$ 
millisecond mean latency irrespective of the cluster size. Meanwhile, Ped-\ourname{}'s
throughput decreases and latency increases as the number of replicas in the cluster
increases. We see that KZG-\ourname{} asympotically is better than Ped-\ourname{}, but
Ped-\ourname{}'s cheaper cryptographic operations still causes it to outperform KZG-\ourname{}.

\subsubsection{\vssReconstructNew Microbenchmark}

\begin{figure}
\begin{subfigure}[t]{0.475\columnwidth}
  \includegraphics[width=1\linewidth]{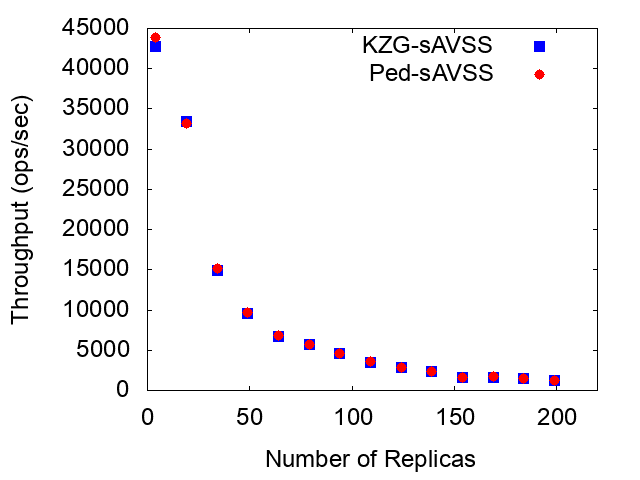}
  \caption{\vssReconstructNew throughput}
  \label{fig:sharereconstruct:xput}
\end{subfigure}
\hfill
\begin{subfigure}[t]{0.475\columnwidth}
  \includegraphics[width=1\linewidth]{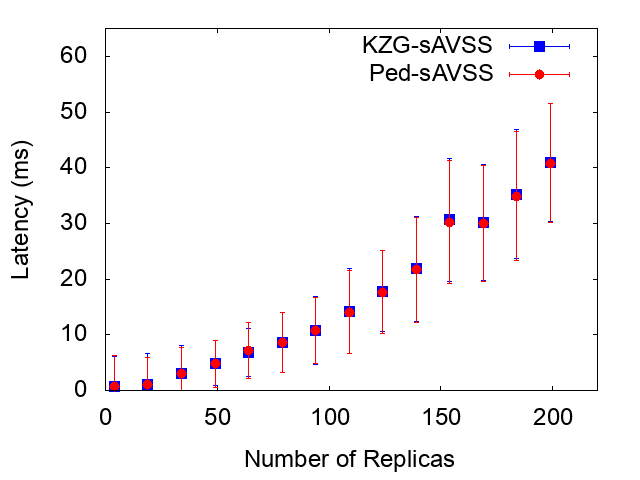}
  \caption{\vssReconstructNew latency}
  \label{fig:sharereconstruct:latency}
\end{subfigure}
\caption{\vssReconstructNew performance with varying \numreplicas}
\label{fig:sharereconstruct}
\end{figure}

\vssReconstructNew has almost identical performance between
KZG-\ourname{} and Ped-\ourname{}, as we see in
\figref{fig:sharereconstruct}.  \vssReconstructNew does not include the
time taken to run \vssVerifyNew since share verification happens when the
message itself is verified. \figref{fig:sharereconstruct:xput} shows that 
\vssReconstructNew can occur at high throughput, with $45000$ operations 
per second with $4$ replicas. However, \vssReconstructNew's throughput
drops off quadratically as the cluster size increases, only being able to
do $1100$ operations per second with $\numreplicas=211$ replicas.
\figref{fig:sharereconstruct:latency} shows a similar performance story,
with the latency increasing quadratically as the cluster size increases
though even at $\numreplicas=211$ the latency is fairly low at $46$
milliseconds.

The reason we see the quadratic behavior in \vssReconstructNew is
that \vssReconstructNew does a quadratic number of modular
multiplications in a $254$ bit prime field. All multiplicative inverses
are precomputed during setup, which makes the runtime of \vssReconstructNew
very fast even though there is a quadratic dropoff.

\subsubsection{\vssRecoverContribNew Microbenchmark}

\begin{figure}
\begin{subfigure}[t]{0.475\columnwidth}
  \includegraphics[width=1\linewidth]{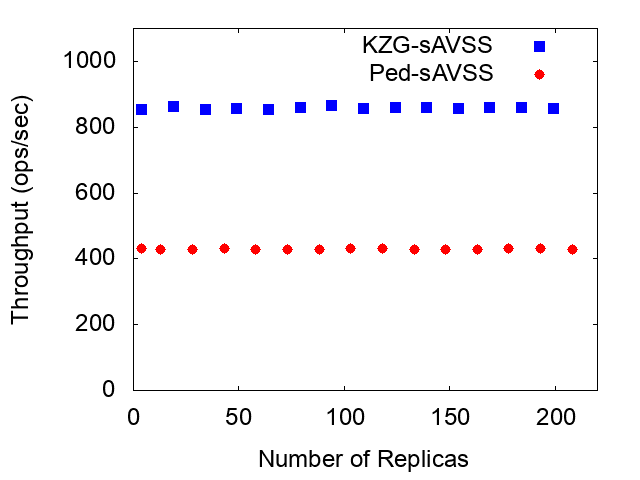}
  \caption{\vssRecoverContribNew throughput}
  \label{fig:sharerecovercontrib:xput}
\end{subfigure}
\hfill
\begin{subfigure}[t]{0.475\columnwidth}
  \includegraphics[width=1\linewidth]{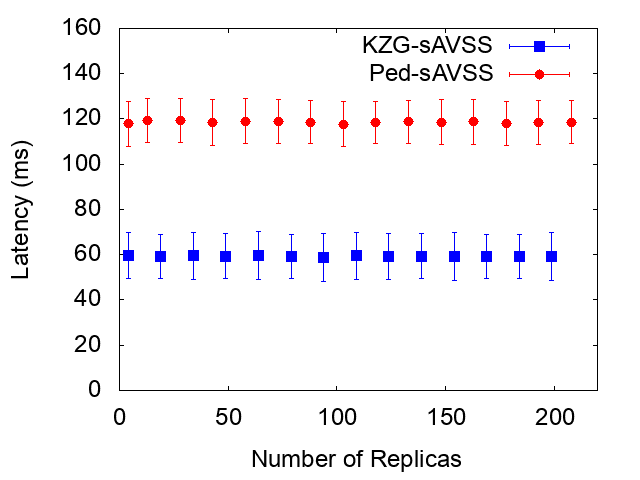}
  \caption{\vssRecoverContribNew latency}
  \label{fig:sharerecovercontrib:latency}
\end{subfigure}
\caption{\vssRecoverContribNew performance with varying \numreplicas}
\label{fig:sharerecovercontrib}
\end{figure}

\figref{fig:sharerecovercontrib} shows that \vssRecoverContribNew throughput and latency 
is independent of the cluster size for both KZG-\ourname{} and Ped-\ourname{}.
Additionally, Ped-\ourname{}'s \vssRecoverContribNew has exactly half the throughput 
($430$ vs $860$) and twice the latency ($118$ ms vs $59$ ms) of KZG-\ourname{}.
The constant CPU cost is due to \vssRecoverContribNew only computing a share of
a DPRF point evaluation and its associated verificaion proof. The reason that Ped-\ourname{}'s
\vssRecoverContribNew operation is exactly half as performant than KZG-\ourname{} is that
Ped-\ourname{} has two polynomial shares that must be recovered while KZG-\ourname{} 
has one. 

\subsubsection{\vssRecoverVerifyNew Microbenchmark}

\begin{figure}
\begin{subfigure}[t]{0.475\columnwidth}
  \includegraphics[width=1\linewidth]{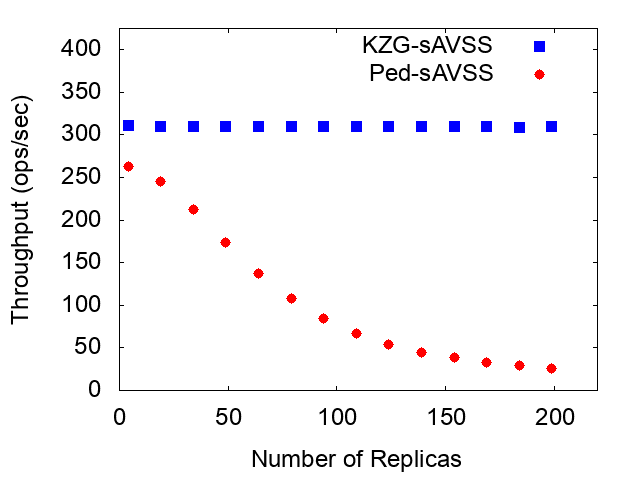}
  \caption{\vssRecoverVerifyNew throughput}
  \label{fig:sharerecoververify:xput}
\end{subfigure}
\hfill
\begin{subfigure}[t]{0.475\columnwidth}
  \includegraphics[width=1\linewidth]{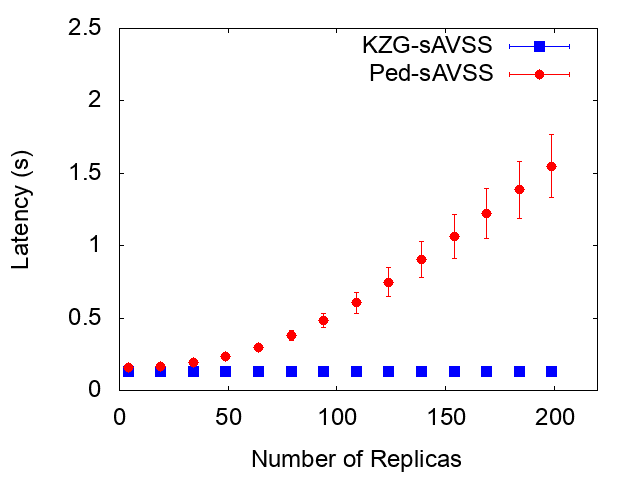}
  \caption{\vssRecoverVerifyNew latency}
  \label{fig:sharerecoververify:latency}
\end{subfigure}
\caption{\vssRecoverVerifyNew performance with varying \numreplicas}
\label{fig:sharerecoververify}
\end{figure}

\figref{fig:sharerecoververify} shows that KZG-\ourname{}'s \vssRecoverVerifyNew
operation has higher throughput and lower latency than Ped-\ourname{}.
KZG-\ourname{} has a $1.18$ times higher throughput at a cluster size of
$\numreplicas=4$, with the performance gap increasing to $13$ times at
larger cluster sizes of $\numreplicas=211$. The difference in latency
is also similar, with KZG-\ourname{}'s \vssRecoverVerifyNew taking $130$ milliseconds
for all cluster sizes while Ped-\ourname{} starts at $150$ milliseconds at
$\numreplicas = 4$ and increases to $1.7$ seconds at $\numreplicas = 211$.

This performance difference occurs since \vssRecoverVerifyNew must combine 
commitments and witnesses from the contributions received from \vssRecoverContribNew.
KZG-\ourname{} performs this computation using a constant number of elliptic curve multiplications
whereas Ped-\ourname{} computes this using a linear number of elliptic curve multiplications.
Thus, as the cluster size increases, Ped-\ourname{}'s performance also degrades
accordingly.

\subsubsection{\vssRecoverNew Microbenchmark}

\begin{figure}
\begin{subfigure}[t]{0.475\columnwidth}
  \includegraphics[width=1\linewidth]{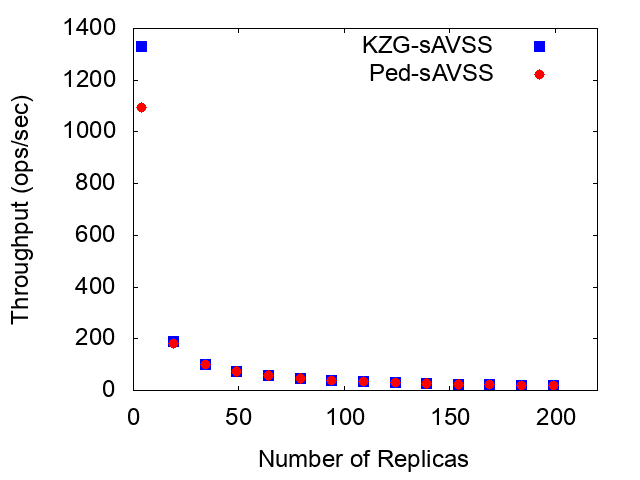}
  \caption{\vssRecoverNew throughput}
  \label{fig:sharerecover:xput}
\end{subfigure}
\hfill
\begin{subfigure}[t]{0.475\columnwidth}
  \includegraphics[width=1\linewidth]{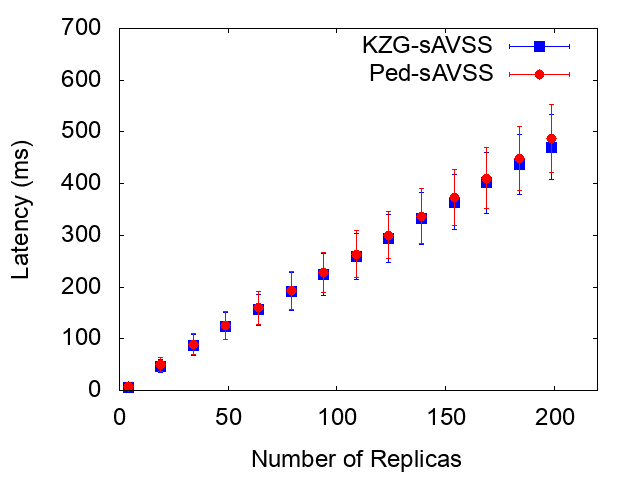}
  \caption{\vssRecoverNew latency}
  \label{fig:sharerecover:latency}
\end{subfigure}
\caption{\vssRecoverNew performance with varying \numreplicas}
\label{fig:sharerecover}
\end{figure}

Similar to \vssReconstructNew, share verification via
\vssRecoverVerifyNew happens in our implementation upon receiving each
share from a replica, and the costs of these \vssRecoverVerifyNew
operations are not included in the \vssReconstructNew results shown in
\figref{fig:sharerecover}.  \vssRecoverNew thus incurs costs primarily
due to interpolation (like \vssReconstructNew), evaluation of the DPRF
and interpolation of any witnesses.  Therefore, asymptotically, we see
in \figref{fig:sharerecover:xput} and
\figref{fig:sharerecover:latency} that \vssRecoverNew behaves
similarly to \vssReconstructNew but with an order of magnitude lower
throughput and an order of magnitude higher latency.

\subsection{Incorporating \ourname{} into PBFT}

We incorporate \ourname{} into a PBFT implementation in order to
implement a threshold trusted third party (T3P).  We instantiate our
T3P using KZG-\ourname{} and Ped-\ourname{}, which we will refer to
as KZG-T3P and Ped-T3P.  We also implement and evaluate a key-value 
store on top of KZG-T3P, Ped-T3P, and PBFT.

To generate load in our evaluation, a client sends PUT requests
asynchronously to the primary. The client pregenerates the requests to 
send to the cluster and loops through them once they are finished. 
For our throughput experiments, the clients asynchronously send enough
requests at a time to saturate the system without oversaturating it.
For our latency benchmarks, the clients send requests serially and
measure the latency of each request.
We used Amazon AWS to run our tests 
and used \texttt{c5.4xlarge} instances for all clients and replicas.

Similar to our microbenchmarks, our implementation uses the
Relic~\cite{relic} cryptographic library for most cryptographic
operations and the \texttt{BN\_P254} elliptic curve. For signatures
in PBFT, our implementation uses the optimized \texttt{secp256k1}
library used in Bitcoin.

\subsubsection{Benchmarks}

\begin{figure}
\begin{subfigure}[t]{0.475\columnwidth}
  \includegraphics[width=1\linewidth]{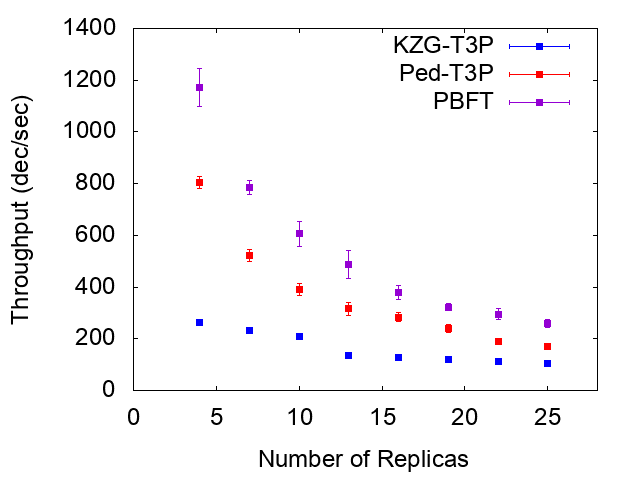}
  \caption{PUT operation throughput}
  \label{fig:e2e:xput}
\end{subfigure}
\hfill
\begin{subfigure}[t]{0.475\columnwidth}
  \includegraphics[width=1\linewidth]{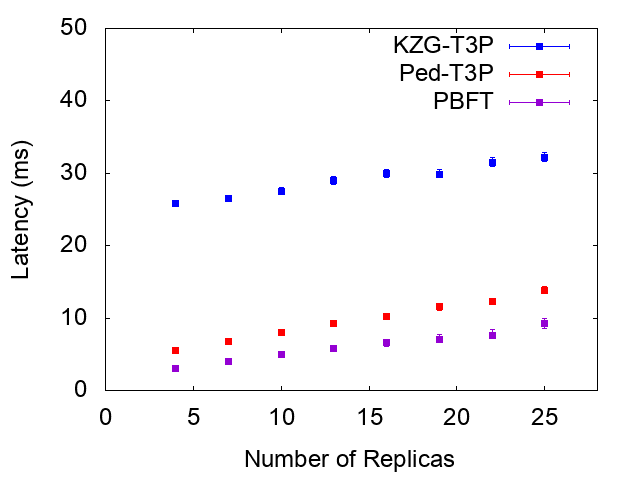}
  \caption{PUT operation latency}
  \label{fig:e2e:latency}
\end{subfigure}
\caption{PUT operation performance with varying \numreplicas}
\label{fig:e2e}
\end{figure}

\figref{fig:e2e:latency} shows that there is some overhead for secret
sharing, as we would expect. However, for all schemes and cluster sizes, 
the latency is less than $35$ milliseconds. While KZG-T3P is slower
than Ped-T3P and PBFT by a factor of $3$-$8$ times, the total latencies are very 
small so the user perceived lag would be insignificant.

\figref{fig:e2e:xput} illustrates the throughput overhead of secret
sharing. At $\numreplicas = 4$, PBFT has
a throughput of $1170$ decisions per second, while Ped-T3P can make
$800$ decisions per second. Thus, we pay a performance penalty of
$32\%$ in order to get secret sharing in a small cluster. The performance
penalty stays roughly constant between Ped-T3P and PBFT. However, since
KZG-T3P is linear, we see the performance penalty shrinking slightly as
the cluster size increases from a factor of $78\%$ at $\numreplicas=4$ to
$60\%$ at $\numreplicas=25$.

Next, we focus on the scalability of the schemes themselves by looking
at the performance dropoff of the system as the cluster size increases.
\figref{fig:e2e:xput} also shows the benefits of a fully linear secret
sharing scheme as well since the performance dropoff of KZG-T3P is
the smallest as the cluster size increased. PBFT took a $78\%$
performance hit in throughput when going from $\numreplicas=4$ and
serving $1170$ decicions per second to $\numreplicas=25$ serving $260$
decisions per second. Similarly, Ped-T3P took a $79\%$ performance hit
as well. However, KZG-T3P, whose performance was dominated by the
expensive secret sharing cost rather than the underlying PBFT
algorithm had a smaller performance dropoff of $60\%$.

\section{Related Work}

This paper makes two primary contributions: an asynchronous verifiable secret sharing scheme 
\ourname{} that has linear dealer cost and a threshold trusted third party (T3P) built by combining
\ourname{} with a Byzantine Fault Tolerant state machine. We discuss the related works in both areas below.

\subsection{Proactive secret sharing}

A treatment of the prior work in verifiable secret sharing and asynchronous verifiable secret
  sharing is given in \secref{sec:overview:existing}.
Unlike those prior works though, another way to approach share recovery is through proactive
  secret sharing.
Using proactive secret sharing for share recovery would require sending a random polynomial that
  has nothing in common with the original shared polynomial except for the share that the recovering
  replica is interested in.

Prior work in proactive secret sharing~\cite{herzberg1995proactive} is difficult to apply directly
  to the problem of share recovery, however.
These works~\cite{herzberg1995proactive, herzberg1997proactive} assume a synchronous broadcast
  channel that delivers to all replicas instantaneously, which greatly simplifies the problem of 
  agreeing on a random recovery polynomial.
PVSS~\cite{zhou2005apss} does not make any such assumption and can be used in \ourname{}, but it 
  suffers from an exponential setup cost in the number of faults it tolerates, making it unusable for 
  tolerating more than a few faults.
MPSS~\cite{schultz2010mpss} uses a Byzantine agreement protocol in order to explicitly agree on
  the random recovery polynomial, which would add a few additional rounds to \ourname{} if used in 
  share reconstruction.
Although closest in "spirit" to proactive recovery schemes, \ourname{} addresses only
the share phase. It is left for future work to see whether proactive share
recovery can be expedited with techniques borrowing from \ourname{}.

\subsection{Privacy in BFT}

Methods to store data across \numreplicas storage nodes in a way
that ensures the privacy, integrity, and availability of the data
despite up to \thresh of these nodes being compromised is a theme that
has been revisited numerous times in the last 30 years
(e.g.,~\cite{herlihy1987how, tompa1988how, deswarte1991intrusion,
  krawczyk1993secret, ganger2000survivable}).  The proposals in this
vein of research often do not defend against the misbehavior of the
data writers.  In particular, a data writer might deploy data to the
storage nodes in a way that makes data recovery impossible or
ambiguious, in the sense that the data reconstructed depends on which
correct nodes cooperate to do so.  Protecting against corrupt data
writers is one of the primary goals of \textit{verifiable} secret
sharing and its derivatives, for which we've surveyed the most
directly related works in \secref{sec:overview}.

With the rise of blockchains supporting smart contracts, there has
been a resurgence of activity in finding ways to add privacy
guarantees to Byzantine fault-tolerant algorithms, and indeed this is
one motivation behind our work.  Another class of approaches to this
problem uses zero knowledge
proofs~\cite{miers2013zerocoin,sasson2014zerocash} for privacy. These
approaches provide a very strong guarantee where it is impossible for
anyone (other than the data owner) to recover the sensitive data, but
where anyone can validate that the data satisfies some prespecified
properties.  However, such systems only work for a limited set of
applications, rather than general purpose state machines that we
target here.  Additionally, these systems do not have any control over
the data itself; i.e., the sensitive data must be managed by the
owner, which is not suitable for a large class of applications.

CALYPSO~\cite{eleftherios2018calypso} resolves this through the use of
a publicly verifiable secret sharing scheme, but they require two BFT
clusters---one for access control and one for secret management.
Thus, their protocol requires more replicas to operate.  Additionally,
CALYPSO requires the access-control policy to be specified ahead of
time by the client, whereas \ourname{} can easily allow dynamic
access-control policies.


\section{Conclusion}
This paper introduces a new method for creating asynchronous verifiable secret sharing (AVSS) schemes for use in Byzantine
Fault Tolerance(BFT) protocols called \ourname{}.
We apply this method to two VSS schemes and incorporate the resulting AVSS schemes into PBFT.
We then implement a Byzantine Fault Tolerant key value store and evaluate the effectiveness of our scheme.

\newpage
\bibliographystyle{IEEEtran}
\bibliography{IEEEabrv,conferences,main}

\begin{thebibliography}{10}
\providecommand{\url}[1]{#1}
\csname url@samestyle\endcsname
\providecommand{\newblock}{\relax}
\providecommand{\bibinfo}[2]{#2}
\providecommand{\BIBentrySTDinterwordspacing}{\spaceskip=0pt\relax}
\providecommand{\BIBentryALTinterwordstretchfactor}{4}
\providecommand{\BIBentryALTinterwordspacing}{\spaceskip=\fontdimen2\font plus
\BIBentryALTinterwordstretchfactor\fontdimen3\font minus
  \fontdimen4\font\relax}
\providecommand{\BIBforeignlanguage}[2]{{%
\expandafter\ifx\csname l@#1\endcsname\relax
\typeout{** WARNING: IEEEtran.bst: No hyphenation pattern has been}%
\typeout{** loaded for the language `#1'. Using the pattern for}%
\typeout{** the default language instead.}%
\else
\language=\csname l@#1\endcsname
\fi
#2}}
\providecommand{\BIBdecl}{\relax}
\BIBdecl

\bibitem{kate2010constant-size}
A.~Kate, G.~M. Zaverucha, and I.~Goldberg, ``Constant-size commitments to
  polynomials and their applications,'' in \emph{Advances in Cryptology --
  ASIACRYPT 2010}, ser. LNCS, vol. 6477, Dec. 2010, pp. 177--194.

\bibitem{damgard2012mpc}
I.~Damg\aa{}rd, V.~Pastro, N.~Smart, and S.~Zakarias, ``Multiparty computation
  from somewhat homomorphic encryption,'' in \emph{Advances in Cryptology --
  CRYPTO 2012}, ser. LNCS, vol. 7417, 2012, pp. 643--662.

\bibitem{nordholt2018minimising}
P.~S. Nordhold and M.~Veeningen, ``Minimising communication in honest-majority
  {MPC} by batchwise multiplication verification,'' in \emph{International
  Conference on Applied Cryptography and Network Security (ACNS) 2018}, ser.
  LNCS, vol. 10892, 2018, pp. 321--339.

\bibitem{chor1985verifiable}
B.~Chor, S.~Goldwasser, S.~Micali, and B.~Awerbuch, ``Verifiable secret sharing
  and achieving simultaneity in the presence of faults,'' in
  \emph{26\textsuperscript{th} IEEE Symposium on Foundations of Computer
  Science (FOCS)}, Oct. 1985, pp. 383--395.

\bibitem{pedersen1991non-interactive}
T.~P. Pedersen, ``Non-interactive and information-theoretic secure verifiable
  secret sharing,'' in \emph{Advances in Cryptology -- CRYPTO '91}, ser. LNCS,
  vol. 576, 1992, pp. 129--140.

\bibitem{castro2002practical}
M.~Castro and B.~Liskov, ``Practical byzantine fault tolerance and proactive
  recovery,'' \emph{ACM Transactions on Computer Systems}, vol.~20, no.~4,
  2002.

\bibitem{kotla2008zyzzyva}
R.~Kotla, A.~Clement, E.~Wong, L.~Alvisi, and M.~Dahlin, ``Zyzzyva: speculative
  byzantine fault tolerance,'' \emph{Communications of the {ACM}}, vol.~51,
  no.~11, pp. 86--95, 2008.

\bibitem{golan-gueta2018sbft}
G.~G. Gueta, I.~Abraham, S.~Grossman, D.~Malkhi, B.~Pinkas, M.~K. Reiter,
  D.~Seredinschi, O.~Tamir, and A.~Tomescu, ``{SBFT}: a scalable decentralized
  trust infrastructure for blockchains,'' \emph{CoRR}, vol. abs/1804.01626,
  2018.

\bibitem{shamir1979how}
A.~Shamir, ``How to share a secret,'' \emph{Communications of the {ACM}},
  vol.~22, no.~11, pp. 612--613, Nov. 1979.

\bibitem{feldman1987practical}
P.~Feldman, ``A practical scheme for non-interactive verifiable secret
  sharing,'' in \emph{28\textsuperscript{th} IEEE Symposium on Foundations of
  Computer Science (FOCS)}, Oct. 1987, pp. 427--438.

\bibitem{cascudo2017scrape}
I.~Cascudo and B.~David, ``{SCRAPE}: Scalable randomness attested by public
  entities,'' in \emph{International Conference on Applied Cryptography and
  Network Security (ACNS)}.\hskip 1em plus 0.5em minus 0.4em\relax Springer,
  2017, pp. 537--556.

\bibitem{canetti1993fast}
R.~Canetti and T.~Rabin, ``Fast asynchronous {Byzantine} agreement with optimal
  resilience,'' in \emph{25\textsuperscript{th}ACM Symposium on Theory of
  Computing (STOC)}, May 1993, pp. 42--51.

\bibitem{cachin2002asynchronous}
C.~Cachin, K.~Kursawe, A.~Lysyanskaya, and R.~Strobl, ``Asynchronous verifiable
  secret sharing and proactive cryptosystems,'' in \emph{9\textsuperscript{th}
  ACM Conference on Computer and Communications Security (CCS)}, Nov. 2002.

\bibitem{bracha1985asynchronous}
G.~Bracha and S.~Toueg, ``Asynchronous consensus and broadcast protocols,''
  \emph{Journal of the ACM}, vol.~32, no.~4, pp. 824--840, Oct. 1985.

\bibitem{backes2013asynchronous}
M.~Backes, A.~Datta, and A.~Kate, ``Asynchronous computational {VSS} with
  reduced communication complexity,'' in \emph{Topics in Cryptology -- CT-RSA
  2013}, ser. LNCS, vol. 7779, Feb. 2013, pp. 259--276.

\bibitem{herzberg1995proactive}
A.~Herzberg, S.~Jarecki, H.~Krawczyk, and M.~Yung, ``Proactive secret sharing
  or: How to cope with perpetual leakage,'' in \emph{Advances in Cryptology --
  CRYPTO '95}, vol. 963, 1995, pp. 339--352.

\bibitem{naor1999distributed}
M.~Naor, B.~Pinkas, and O.~Reingold, ``Distributed pseudo-random functions and
  {KDCs},'' in \emph{Advances in Cryptology -- EUROCRYPT '99}, ser. LNCS, vol.
  1592, 1999, pp. 327--346.

\bibitem{cachin2005constantinople}
C.~Cachin, K.~Kursawe, and V.~Shoup, ``Random oracles in {Constantinople}:
  Practical asynchronous {Byzantine} agreement using cryptography,''
  \emph{Journal of Cryptology}, vol.~18, pp. 219--246, Jul. 2005.

\bibitem{miller2016honeybadger}
A.~Miller, Y.~Xia, K.~Croman, E.~Shi, and D.~Song, ``The honey badger of {BFT}
  protocols,'' in \emph{23\textsuperscript{rd} ACM Conference on Computer and
  Communications Security (CCS)}, Oct. 2016, pp. 31--42.

\bibitem{herlihy1990linearizability}
M.~P. Herlihy and J.~M. Wing, ``Linearizability: A correctness condition for
  concurrent objects,'' \emph{ACM Transactions on Programming Languages and
  Systems}, vol.~12, pp. 463--492, Jul. 1990.

\bibitem{dwork1988consensus}
C.~Dwork, N.~Lynch, and L.~Stockmeyer, ``Consensus in the presence of partial
  synchrony,'' \emph{Journal of the ACM}, vol.~35, no.~2, pp. 288--323, Apr.
  1988.

\bibitem{martin2006fab}
J.-P. Martin and L.~Alvisi, ``Fast {Byzantine} consensus,'' \emph{IEEE
  Transactions on Dependable and Secure Computing}, vol.~3, no.~3, pp.
  202--215, Jul. 2006.

\bibitem{kotla2009zyzzyva}
R.~Kotla, L.~Alvisi, M.~Dahlin, A.~Clement, and E.~L. Wong, ``Zyzzyva:
  Speculative {Byzantine} fault tolerance,'' \emph{ACM Transactions on Computer
  Systems}, vol.~27, no.~4, 2009.

\bibitem{relic}
D.~F. Aranha and C.~P.~L. Gouv\^{e}a, ``{RELIC is an Efficient LIbrary for
  Cryptography},'' \url{https://github.com/relic-toolkit/relic}.

\bibitem{herzberg1997proactive}
A.~Herzberg, M.~Jakobsson, S.~Jarecki, H.~Krawczyk, and M.~Yung, ``Proactive
  public key and signature systems,'' in \emph{4\textsuperscript{th}ACM
  Conference on Computer and Communications Security (CCS)}, Apr. 1997, pp.
  100--110.

\bibitem{zhou2005apss}
L.~Zhou and F.~B. Schneider, ``{APSS}: Proactive secret sharing in asynchronous
  systems,'' \emph{ACM Transactions on Information and System Security},
  vol.~8, no.~3, pp. 259--286, Aug. 2005.

\bibitem{schultz2010mpss}
D.~Schultz, B.~Liskov, and M.~Liskov, ``{MPSS}: Mobile proactive secret
  sharing,'' \emph{ACM Transactions on Information and System Security},
  vol.~13, no.~4, pp. 1--32, Dec. 2010.

\bibitem{herlihy1987how}
M.~P. Herlihy and J.~D. Tygar, ``How to make replicated data secure,'' in
  \emph{Advances in Cryptology -- CRYPTO '87}, ser. LNCS, vol. 293, 1988, pp.
  379--391.

\bibitem{tompa1988how}
M.~Tompa and H.~Woll, ``How to share a secret with cheaters,'' \emph{Journal of
  Cryptology}, vol.~1, pp. 133--138, 1988.

\bibitem{deswarte1991intrusion}
Y.~Deswarte, L.~Blain, and J.-C. Fabre, ``Intrusion tolerance in distributed
  computing systems,'' in \emph{IEEE Symposium on Security and Privacy}, May
  1991, pp. 110--121.

\bibitem{krawczyk1993secret}
H.~Krawczyk, ``Secret sharing made short,'' in \emph{Advances in Cryptology --
  CRYPTO '93}, ser. LNCS, vol. 773, 1994, pp. 136--146.

\bibitem{ganger2000survivable}
G.~R. Ganger, P.~K. Khosla, M.~Bakkaloglu, M.~W. Bigrigg, G.~R. Goodson,
  S.~Oguz, V.~Pandurangan, C.~A.~N. Soules, J.~D. Strunk, and J.~J. Wylie,
  ``Survivable storage systems,'' \emph{IEEE Computer}, vol.~33, pp. 61--68,
  2000.

\bibitem{miers2013zerocoin}
I.~Miers, C.~Garman, M.~Green, and A.~D. Rubin, ``{Zerocoin}: Anonymous
  distributed e-cash from bitcoin,'' in \emph{IEEE Symposium on Security and
  Privacy 2013}, 2013, pp. 397--411.

\bibitem{sasson2014zerocash}
E.~B. Sasson, A.~Chiesa, C.~Garman, M.~Green, I.~Miers, E.~Tromer, and
  M.~Virza, ``{Zerocash}: Decentralized anonymous payments from bitcoin,'' in
  \emph{IEEE Symposium on Security and Privacy 2014}, 2014, pp. 459--474.

\bibitem{eleftherios2018calypso}
E.~Kokoris-Kogias, E.~C. Alp, S.~D. Siby, N.~Gailly, L.~Gasser, P.~Jovanovic,
  E.~Syta, and B.~Ford, ``{CALYPSO}: Auditable sharing of private data over
  blockchains,'' Cryptology ePrint Archive, Report 2018/209, 2018,
  \url{https://eprint.iacr.org/2018/209}.

\end{thebibliography}

\newpage
\appendices
\section{Security}
\label{sec:app:pbftproof}

In this appendix, we show why our composition of \ourname{} along with PBFT is secure.
We do this by first observing under what conditions the share recovery protocol will terminate.
To show linearizability and liveness, we map every execution of our modified PBFT algorithm
  to the original PBFT algorithm.
Thus, since the original PBFT algorithm satisfies linearizability and liveness, so does our
  modified algorithm.
Then, we show privacy separately.

\paragraph{Share Recovery Protocol Termination.}
We claim that the share recovery protocol will always terminate if $\numfaulty + 1$ replicas
  have successfully completed the sharing and the network eventually delivers all messages.
To see why, recall that a replica that is missing its share needs the output of 
  \vssRecoverContribNew from $\numfaulty + 1$ replicas.
If $\numfaulty + 1$ replicas are honest, then they will faithfully call \vssRecoverContribNew
  and send the output to a replica that is missing its share.
The missing share can then be recovered by using \vssRecoverNew to terminate the share recovery
  protocol.

\paragraph{Normal Case Protocol}
In the normal case protocol, we only have changed how the client constructs requests.
If a client is honest, then we can simply ignore the secrets being shared in the request and
  have the client send regular requests in the original run of the PBFT algorithm.
The requests are consistent due to the binding property of our verifiable secret sharing (VSS)
  scheme.
If a client is dishonest and sends an invalid share to the replica, then in the original PBFT
  protocol run the client will drop the request message to the replica.
Now, when the message is dropped, the replica can still obtain the request from another replica
  in the system.
In the modified protocol, this is done using the share recovery protocol so we simply wait to 
  deliver the request messages in the original run until the share recovery protocol terminates.
Note that if the share recovery protocol never terminates in the modified protocol, that means
  that less than $\numfaulty + 1$ honest replicas have the request.
This means that strictly less than $2\numfaulty + 1$ total replicas have the request, making it
  impossible for this request to be prepared.
Therefore, if the modified normal case protocol never terminates, then neither does the original
  protocol.
Thus, we see that liveness is unchanged from the original PBFT protocol.

Additionally, through the binding property of our underlying VSS scheme, we know that if a request
  has been committed, all secret values must be consistently shared.
Thus, we see that the linearizability property also follows from linearizability in the original
  protocol along with binding.

\paragraph{Checkpoint protocol}
The checkpoint protocol is idential to a case where the state of the replicated service contains
  only the commitments of the secret values instead of the secret values themselves.
Thus, by the binding property of the underlying VSS scheme, we have a one to one mapping from a
  run of the checkpoint protocol for our modified PBFT algorithm and the original PBFT algorithm.
Therefore, if the original PBFT's checkpoint algorithm provides liveness and linearizability, then
  so does our modified algorithm.

\paragraph{State Transfer Protocol}
The state transfer protocol can be mapped back similarly to the normal case protocol.
A replica receiving the value of a key using the share recovery protocol in our modified PBFT
  would have been receiving the plaintext value of the key in the original PBFT algorithm.
We simply delay the plaintext value of the key until the share recovery protocol completes in
  our modified PBFT protocol.
Additionally, in the state transfer protocol, we know that the share recovery protocol will
  complete since at least $2\numfaulty + 1$ replicas have the state at the last checkpoint.
This means that at least $\numfaulty + 1$ honest replicas have the state, which is sufficient
  to guarantee termination during periods of synchrony.

\paragraph{Privacy}
Our modified PBFT protocol achieves privacy through the hiding property of the VSS protocol.
The hiding property says that a legitimate adversary (i.e. one that has at most $\numfaulty$ 
  shares of the secret) cannot do nonnegligibly better than guessing the secret at random.
Thus, privacy is satisfied unless an adversay gets at least $\numfaulty + 1$ shares of a value.
However, this means that some correct replica has shared the secret with the adversary which
  contradicts our threshold assumption.
Thus, we see that our modified PBFT protocol preserves linearizability and liveness while also
  guaranteeing privacy.

\section{PBFT State Transfer and View Change}
\label{sec:app:protocoldescription}

In this section, we describe the PBFT state transfer and view change protocols.

\subsection{View Change Protocol} 

The view change protocol changes the leader. 
The core mechanism for transferring safe values across views is
for a new leader to collect a set $P$ of
view-change messages from a quorum of $2f+1$ replicas. Each replica sends a
view-change message containing the replica's \textit{local state}:
Its local request-log,
and the commit-certificate with the highest view number it responded to with a
commit message, if any.

The leader processes the set $P$ as follows.

\begin{enumerate}
 \item
  Initially, it sets a \emph{leader-log} $G$ to an empty log.
  
  \item
  If any view-change message contains a valid commit-certificate, then it
selects the one with the highest view number and copies its log to $G$.
Share recovery is triggered for any requests in $G$ that the leader is missing
its private share.
 
   
\end{enumerate}

The leader sends a new-view message to all replicas. The message
includes the new view number, the set $P$ of view-change messages the
leader collected as a \emph{leader-proof} for the new view, and the \emph{leader-log}
$G$. A replica accepts a new-view
message if it is valid, and \emph{adopts} the leader log. It may need to roll
back speculatively executed requests, and process new ones. As usual, processing
may entail triggering share-recovery for any requests where the replica is
missing its private share. 


\subsection{State Transfer Protocol}
We present a modified version of the PBFT state transfer protocol that is simpler and more
  suited when TCP is used for the underlying network protocol.
When a replica has fallen behind, it sends a \emph{state transfer request} along with its
  current sequence number to at least $\numfaulty + 1$ replicas.
Some replica will respond with the most recent valid checkpoint messages and the messages
  from the normal case protocol that were missed by the slow replica.
In addition, the response will contain only the values of the keys that have
  changed since the sequence number known to the slow replica as well as the full requests
  that came after the last checkpoint.

\end{document}